\documentclass[a4paper,fleqn,usenatbib]{mnras}

\usepackage[T1]{fontenc}
\usepackage{ae,aecompl}
\usepackage{ulem,soul}

\usepackage{graphicx}	
\usepackage{amsmath}	
\usepackage{amssymb}	

\usepackage{graphicx,color}
\usepackage[usenames,dvipsnames,svgnames,table]{xcolor}

\title[Secular Gravitational Instability]{Linear Analysis of  the Nonaxisymmetric Secular Gravitational Instability}
\author[M. Shadmehri, R. Oudi, G. Rastegarzadeh]{
Mohsen Shadmehri $^1$\thanks{E-mail: m.shadmehri@gu.ac.ir},
Razieh Oudi $^{2}$, Gohar Rastegarzadeh $^{2}$
\\
$^1$Department of Physics, Faculty of Sciences, Golestan University, Gorgan 49138-15739, Iran\\
$^{2}$ Department of Physics, Semnan University, Semnan 35196-45399, Iran\\
}

\date{Accepted XXX. Received YYY; in original form ZZZ}

\pubyear{2017}

\begin{document}
\label{firstpage}
\pagerange{\pageref{firstpage}--\pageref{lastpage}}
\maketitle

\begin{abstract}
In protoplanetary discs (PPDs) consisting of gas and dust particles,  fluid instabilities induced by the drag force, including secular gravitational instability (SGI) can facilitate planet formation. Although SGI subject to the axisymmetric perturbations was originally studied in the absence of gas feedback and it then generalized using a two-fluid approach, the fate of the nonaxisymmetric SGI, in either case, is an unexplored problem. We present a linear perturbation analysis of the nonaxisymmetric SGI in a PPD by implementing a two-fluid model. We explore the growth of the local, nonaxisymmetric perturbations using a set of linearized perturbation equations in a sheared frame. The nonaxisymmetric perturbations display a significant growth during a finite time interval even when the system is stable against the axisymmetric perturbations. Furthermore,  the surface density perturbations do not show the continuous growth but are temporally amplified.  We also study cases where the dust component undergoes amplification whereas the gas component remains stable. The amplitude amplification, however, strongly depends on the model parameters.  In the minimum mass solar nebula (MMSN), for instance, the dust fluid amplification at the radial distance 100 au occurs when the Stokes number is about unity. But the amplification factor reduces as the dust and gas coupling becomes weaker. Furthermore, perturbations with a larger azimuthal wavelength exhibit a larger amplification factor. 
\end{abstract}




\section{Introduction}

Understanding planet formation mechanisms in  protoplanetary discs (PPDs)  is still a controversial issue despite considerable achievements  in recent years  \citep{Lissauer2007, Durisen2007, Helled2014}. The most widely studied planet formation theories are the so-called {\it core accretion} model which is efficient in the inner region of a PPD \citep[e.g.,][]{Mizuno1980, Stevenson1982, Pollack1996}  and {\it gravitational instability} (GI)   which operates mainly in the outer regions \citep[e.g.,][]{Adams89,Boss97,Boley2009, Rafikov2009}. However, alternative planet formation theories have been proposed during recent years, including inside-out planet formation scenario \citep{Chatt2014} and tidal downsizing model \citep{Nayakshin2010}. In the core accretion model,  when the mass of an already formed rocky core becomes about $10M_{\oplus}$,    this planetary embryo  is able to grow further through the accretion of its  ambient gas and dust particles onto it \citep{Cameron1973,Hayashi1977,Lissauer1993,D'Angelo2010}.

Linear perturbation analysis and numerical simulations show that a gaseous disc with the surface density $\Sigma$, Keplerian angular velocity $\Omega$ and the sound speed $c_{\rm s}$  is gravitationally unstable subject to the axisymmetric perturbations if Toomre parameter, i.e.,  $Q \equiv c_{\rm s}\Omega/(\pi G \Sigma)$ becomes less than a critical value around unity \citep{Toomre1964}.  This criterion  has been successfully implemented in the star formation theories  in galaxies \citep[e.g.,][]{Collin2008,Krumholz2010,Romeo2014,Goldbaum2016} and planet formation scenarios in the PPDs \citep[e.g.,][]{Matzner2005,Kratter2008,Rafikov2005,Boley2009, Rafikov2009}. \cite{Gammie2001} suggested that Toomre diagnostic is only a necessary condition for disc fragmentation and it does not guarantee to have long-lived fragments in a turbulent self-gravitating disc \citep[also see,][]{Rice2003,Mejia2005,Boss2017}.  But recent numerical simulations of the self-gravitating discs have shown that small enough cooling time scale  is not a sufficient condition for fragmentation \citep{Tsukamoto2015}. \cite{Takahashi2016} showed that formation of the spiral arms in a PPD and their fragmentation are the essential phases of a disc fragmentation. They proposed a revised condition for the fragmentation.

While the primary focus of most disc stability studies is the gas component and its evolution,  about one percent of a PPD total mass is attributed to the dust particles with different sizes  \citep{Natta2007}. Presence of dust particles provides valuable insights about dynamical and chemical structure of a PPD \citep{Vasyunin2011, Akimkin2013, Woitke2016, Rab2017}, its ionization level \citep{Okuzumi2011, Akimkin2015, Ivlev2016, Rab2017}  and radiative transfer through a disc \citep{Akimkin2013,Rab2017}. Formation of rocky planets or cores of gaseous giant planets is  explained based on the  collective instabilities associated with the dust component. In the early works on the clumping of dust particles either purely dusty discs have been considered \citep{Safronov1972,Goldreich1973} or their implemented approximations were rather restrictive \citep{cora,sekiya83,noh91}. 

More recent progress for understanding  mechanisms of dust clumping in a PPD relies on the existence of a relative velocity between gas and dust particles  and the associated drag force.  A dust layer is generally believed to be formed at a PPD midplane due to the sedimentation of these particles. Their rotational velocity is Keplerian, whereas the gas component is rotating with a sub-Keplerian velocity because of the radial gradient of pressure. In a PPD as a mixture of gas and dust, therefore, these components may experience a relative velocity. Since the aerodynamically friction force is directly proportional to this relative velocity, motions of dust particles are  affected by  this force. The gas motion, however,  is unaffected by the dust movement when the dust-to -gas density ratio is small. 

A mechanism of the so-called {\it streaming instability} (SI) is triggered because of the dust movement through the gas \citep{youdin2005,youdin2007,Jac}. However, its efficiency strongly depends on the dust-to-gas density ratio and the dust-gas coupling which is quantified in terms of the Stokes number, i.e., ${\rm St}=t_{\rm stop} \Omega$ where $t_{\rm stop}$ is the stopping time and $\Omega$ is Keplerian angular velocity. For relatively well-coupled particles (${\rm St}<1$), a mechanism  known as {\it secular gravitational instability} (SGI) is driven by the drag force \citep{youdin2011,Cuzzi,Mich}. This mechanism  is  a dissipative version of the classical Toomre analysis for the two-component discs \citep{youdin2011,Cuzzi,Mich,Taka}.  The  SGI has primarily been studied using a  single fluid model where the dust dynamics is treated in a gaseous background with no backreaction of the dust. The onset of SGI, therefore, is found to be unconditional and it is triggered no matter how  dusty layer is thin or thick \citep{youdin2011}. This interesting feature of the SGI is lost when the gas feedback is included in a two-fluid model \citep{Taka}. Numerical simulations of the SGI, however, are needed to address whether this mechanism leads to an appreciable enhancement of the dust surface density \citep{Tominaga2018}. 

Recently observed multiple concentric ringlike structures in the PPDs are commonly interpreted as a sign of the newly born planets at their early formation phase \citep{Andrews2011,Mayama2012,Yen16,Loomis2017,Hendler2017,Dipierro18,van2018}. But we note that  a planet associated with these rings or gaps has not yet been observed directly. Non-planet-related scenarios, thereby, have also been proposed as mechanisms of the multiple ring formation \citep[e.g.,][]{Oku16,Suriano18}.  \cite{TaInu2016} applied their two-fluid model of the SGI to explain HL Tau rings  resulting from this instability. They studied growth  time-scale and  unstable wavelengths   and proposed that SGI is able to create ring structures in the HL Tau disc.  \cite{Latter17}, on the other hand, suggested that these ring-like structures probably unrelated to the SGI. In an alternative mechanism \citep{Oku16}, however, ring-like feature in the HL Tau disc is explained by incorporating sintering in a dust growth model.

Although the focus of the recent studies is to provide an explanation for the ring-like structures, some discs also exhibit complex non-axisymmetric patterns. Spiral arm-like structures, for instance, have been observed in the PPDs such as MWC 7588  \citep{Grady2013,Benisty2015} and SAO 206462 \citep{Muto2012,Garufi2013}. Spiral arms that result from non-axisymmetric perturbations are observed  in near infrared scattered light \citep{Muto2012,Wagner2015} and also in sub-millimeter \citep{Tobin2016}.

These spiral features are commonly studied in terms of planet-disc interactions \citep{Zhu2015,Lee2016,Bae2017}. In this scenario, spiral waves are generated due to the interaction of a planer with  epicyclic oscillations in a disc. A set of these waves, which are in phase, create constructive interference and can create spiral arms \citep{Ogilvie2002,Zhu2015,Lee2016}.

In an alternative explanation, however, the spiral arms are generated due to the non-axisymmetric development of the gravitational instability   \citep{Rice2003,Dong2015,Tomida2017,Juhasz2018}. Spiral arm formation in the galactic scale is a well-studied problem known as spiral density wave theory \citep{Lin1964,Julian1966,Toomre1981}, and we think that a similar approach can be implemented in scenarios of GI-induced structures  in PPDs.  In this regard, SGI provides a natural route of ring-like patterns in a PPD. But prior linear studies of SGI  are restricted  to only  axisymmetric  perturbations. Therefore,  the recent observed non-axisymmetric features in PPDs motivates us to  investigate SGI subject to the non-axisymmetric perturbations.

In this paper, we generalize SGI to a case with nonaxisymmetric perturbations in a disc composed of the gas and dust particles.      Our sheared two-fluid  disc model and the main equations are presented in Section 2. We then present a linear analysis and a set of ordinary differential equations are obtained for the evolution of the perturbations in Section 3. These equations are solved numerically in Section 4. We  conclude with a summary of our main results in Section 5.

\section{Basic Equations}
We persist with already studied SGI models and their general formulation \cite[e.g.,][]{Taka}, however, our focus is to explore  fate of the non-axisymmetric perturbations. A shearing sheet model  \citep{gold}, as a representation  of a razor thin disc small portion, is constructed where its center is at a fixed radial distance $r_0$. The $x$-axis is defined in the radial direction and the $y$-axis is oriented in the azimuthal direction. Thus, position of any point in this rotating plane is $(x,y)=(r-r_{\rm 0},r_{\rm 0}(\theta -\Omega t))$, where $\Omega =(GM/r_{0}^3)^{1/2}$ is the Keplerian angular velocity at $r=r_{0}$. Here, $ M$ is the star mass. In our analysis we use the relation $\Omega=-\Omega_{0}= B- A$, where $A$ and $B$ are the standard Oort constants and ${\bf \Omega}=\Omega {\bf z}$ is the Keplerian angular velocity.

Our basic equations, therefore, are the  continuity, momentum,  and the Poisson equations for the dust fluid and gas component:

\begin{equation}
\frac{\partial\Sigma_{\rm g}}{\partial t}+{\bf \nabla}.(\Sigma_{\rm g} {\bf V}_{\rm g})=0,
\end{equation}

\begin{align}
\Sigma_{\rm g} \left (\frac{\partial {\bf V}_{\rm g}}{\partial t}+({\bf V}_{\rm g}.{\bf\nabla}){\bf V}_{\rm g}+2{\bf \Omega}\times {\bf V}_{\rm g}-\Omega^2 {\bf r} \right )=-c_{\rm s}^2{\bf \nabla}\Sigma_{\rm g}
\notag \\
-\Sigma_{\rm g}{\bf \nabla}(\psi_{\rm g}+\psi_{\rm d})+\frac{\Sigma_{\rm d}({\bf V}_{\rm d}-{\bf V}_{\rm g})}{t_{\rm stop}}
\notag\\
+\frac{\partial}{\partial x_{\rm k}}[\Sigma_{\rm g}\nu(\frac{\partial v_{\rm i g}}{\partial {x}_{\rm k}}+\frac{\partial v_{\rm kg}}{\partial {x}_{\rm i}}-\frac{2}{3}\delta _{\rm ik}\frac{\partial v_{\rm lg}}{\partial {x}_{\rm l}})],
\end{align}

\begin{equation}
\frac{\partial\Sigma_{\rm d}}{\partial t}+{\bf \nabla}.(\Sigma_{\rm d} {\bf V}_{\rm d})=D \nabla^2\Sigma_{\rm d},
\end{equation}

\begin{align}
\Sigma_{\rm d} \left (\frac{\partial {\bf V}_{\rm d}}{\partial t}+({\bf V}_{\rm d}.{\bf\nabla}){\bf V}_{\rm d}+2{\bf \Omega}\times {\bf V}_{\rm d}-\Omega^2 {\bf r} \right )=-c_{\rm d}^2{\bf \nabla}\Sigma_{\rm d}
\notag \\
-\Sigma_{\rm d}{\bf \nabla}(\psi_{\rm g}+\psi_{\rm d})+\frac{\Sigma_{\rm d}({\bf V}_{\rm g}-{\bf V}_{\rm d})}{t_{\rm stop}},
\end{align}

\begin{equation}
\nabla^2(\psi_{\rm g}+\psi_{\rm d})=4\pi G (\Sigma_{\rm g}+\Sigma_{\rm d})\delta(z).
\end{equation}
The subscripts "g" and "d" stand for the gas and dust components. Thus, $\Sigma_{\rm g}$ and $\Sigma_{\rm d}$ are the gas and the dust surface densities. The velocities of the gas and the dust fluids are denoted by $\bf V_{\rm g}$ and $\bf V_{\rm d}$. Furthermore, $\psi_{\rm g}$ and $\psi_{\rm d}$ represent the gravitational potential associated with the gas and dust components. The gas sound speed and the velocity dispersion of the dust particles are $c_{\rm s}$ and $c_{\rm d}$ respectively. Here, $\nu$ is the kinematic viscosity which is written as $\nu=\alpha c_{\rm s}^2 \Omega^{-1}$, where $\alpha$ is the dimensionless measure of the turbulent strength.  Finally, $D$ is the diffusivity of the dust due to the gas turbulence.

In order to close the equations of the system, a relation between $c_{\rm s}$ and $c_{\rm d}$ and an equation for the diffusivity coefficient $D$ are needed.  \cite{youdin2007}  found the following relation:
\begin{equation}
c_{\rm d}^2=\alpha c_{\rm s}^2[\frac{1+2({\rm St})^2+\frac{5}{4}({\rm St})^3}{(1+({\rm St})^2)^2}]
\end{equation}
where ${\rm St}$ is the Stokes number, i.e. ${\rm St}=t_{\rm stop}\Omega_{0}$. Furthermore, the diffusivity of the dust $D$  is written as $ D=\xi c_{\rm s}^2\Omega_{0}^{-1}$, where $\xi$ is the dimensionless diffusion coefficient and is defined as
\begin{equation}
\xi=\alpha[\frac{1+({\rm St})+4({\rm St})^2}{(1+({\rm St})^2)^2}].
\end{equation}
The Stokes number depends on the particle size, $a$, and the mean free path of the molecules, $\lambda$. When the size of particles is smaller than the mean free path of the molecules, i.e. $a<9\lambda /4$, which is known as Epstein regime, the Stokes number is written as ${\rm St}=({\rho_{\rm m} a}/{\rho_{g} v_{\rm th}}) f_{\rm d}^{-1} \Omega$, where $\rho_{g}$ is the gas density and $\rho_{\rm m}=2 $ g cm$^{-3}$ denotes the homogeneous material  density of a dust particle. We also have $v_{\rm th}=(8/\pi)^{1/2} c_{\rm s}$ and $f_{d}=[1+({9 \pi}/{128}) ({\left\| \Delta v \right\|}/{c_{\rm s}})^2]^{1/2} \simeq 1$, where the relative velocity between dust and gas, i.e. $\Delta v$, is much smaller than the sound speed at the disc midplane \citep[e.g.,][]{Miyake2016}.

\section{Linear perturbations}
The two-fluid equations admit an equilibrium configuration with  a constant gas surface density $\Sigma_{\rm 0g}$   and a constant dust surface density $\Sigma_{\rm 0d}$. The dust-to-gas density ratio is  defined via $\epsilon=\Sigma_{\rm 0d}/ \Sigma_{\rm 0g}$ which is an input model parameter. The gas and dust components are undergoing Keplerian motion, i.e. ${\bf V}_{\rm 0g}={\bf V}_{\rm 0d}=2Ax{\bf j}$. Therefore, the linearised equations are
\begin{equation}
\frac{\partial(\delta\Sigma_{\rm g})}{\partial t}+2Ax\frac{\partial(\delta\Sigma_{\rm g})}{\partial y}+\Sigma_{\rm 0g}(\frac{\partial(\delta v_{ x{\rm g}})}{\partial x}+\frac{\partial(\delta v_{ y{\rm g}})}{\partial y})=0,
\end{equation}

\begin{align}
\frac{\partial(\delta v_{ x{\rm g}})}{\partial t}+2Ax\frac{\partial(\delta v_{ x{\rm g}})}{\partial y}-2\Omega (\delta v_{y{\rm g}})=-\frac{\partial}{\partial x}(\delta\psi_{\rm g}+\delta\psi_{\rm d})
\notag \\
 -\frac{c_{\rm s}^2}{\Sigma_{\rm 0g}}\frac{\partial(\delta\Sigma_{\rm g})}{\partial x}+\frac{\epsilon(\delta v_{x{\rm d}}-\delta v_{x{\rm g}})}{t_{\rm stop}}
 \notag\\
 +\frac{4}{3}\nu\frac{\partial^2(\delta v_{x{\rm g}})}{\partial x^2}+\nu\frac{\partial^2(\delta v_{x{\rm g}})}{\partial y^2}+\frac{2A\nu}{\Sigma_{\rm 0g}}\frac{\partial(\delta \Sigma_{\rm g})}{\partial y}+\nu \frac{\partial^2}{\partial y \partial x}(\delta v_{y{\rm g}}),
\end{align}

\begin{align}
\frac{\partial(\delta v_{y{\rm g}})}{\partial t}+2Ax\frac{\partial(\delta v_{y{\rm g}})}{\partial y}+2B (\delta v_{x{\rm g}})=-\frac{\partial}{\partial y}(\delta\psi_{\rm g}+\delta\psi_{\rm d})
\notag\\
-\frac{c_{\rm s}^2}{\Sigma_{\rm 0g}}\frac{\partial(\delta\Sigma_{\rm g})}{\partial y}+\frac{\epsilon(\delta v_{y{\rm d}}-\delta v_{y{\rm g}})}{t_{\rm stop}}
\notag\\
+\frac{4}{3}\nu\frac{\partial^2(\delta v_{y{\rm g}})}{\partial y^2}+\nu\frac{\partial^2(\delta v_{y{\rm g}})}{\partial x^2}+\frac{2A\nu}{\Sigma_{\rm 0g}}\frac{\partial(\delta \Sigma_{\rm g})}{\partial x}+\nu \frac{\partial^2}{\partial x \partial y}(\delta v_{x{\rm g}}),
\end{align}

\begin{align}
\frac{\partial(\delta\Sigma_{\rm d})}{\partial t}+2Ax\frac{\partial(\delta \Sigma_{\rm d})}{\partial y}+\Sigma_{\rm 0d}(\frac{\partial(\delta v_{x{\rm d}})}{\partial x}+\frac{\partial(\delta v_{y{\rm d}})}{\partial y})
\notag\\
=D(\frac{\partial^2}{\partial x^2}+\frac{\partial^2}{\partial y^2})\delta\Sigma_{\rm d},
\end{align}

\begin{align}
\frac{\partial(\delta v_{x{\rm d}})}{\partial t}+2Ax\frac{\partial(\delta v_{x{\rm d}})}{\partial y}-2\Omega (\delta v_{y{\rm d}})=-\frac{\partial}{\partial x}(\delta\psi_{\rm g}+\delta\psi_{\rm d})
\notag\\
-\frac{c_{\rm d}^2}{\Sigma_{\rm 0d}}\frac{\partial(\delta\Sigma_{\rm d})}{\partial x}+\frac{(\delta v_{x{\rm g}}-\delta v_{x{\rm d}})}{t_{\rm stop}},
\end{align}

\begin{align}
\frac{\partial(\delta v_{y{\rm d}})}{\partial t}+2Ax\frac{\partial(\delta v_{y{\rm d}})}{\partial y}+2B(\delta v_{x{\rm d}})=-\frac{\partial}{\partial y}(\delta\psi_{\rm g}+\delta\psi_{\rm d})
\notag\\
-\frac{c_{\rm d}^2}{\Sigma_{\rm 0d}}\frac{\partial(\delta\Sigma_{\rm d})}{\partial y}+\frac{(\delta v_{y{\rm g}}-\delta v_{y{\rm d}})}{t_{\rm stop}},
\end{align}

\begin{equation}
(\frac{\partial^2}{\partial x^2}+\frac{\partial^2}{\partial y^2}+\frac{\partial^2}{\partial z^2})(\delta\psi_{\rm g}+\delta\psi_{\rm d})=4\pi G(\delta\Sigma_{\rm g}+\delta\Sigma_{\rm d})\delta(z),
\end{equation}
where $\delta(z)$ is Dirac delta function.

For exploring non-axisymmetric perturbations, it is more convenient to use a shearing sheet coordinates $(x',y',z')$ introduced by \cite{gold}. We have
\begin{equation}
x'=x,\,\,\,\,\,\,\,\,\,y'=y-2Axt,\,\,\,\,\,\,\,z'=z,\,\,\,\,\,\,\,t'=t,
\end{equation}
where
\begin{align}
\frac{\partial}{\partial x}\equiv\frac{\partial}{\partial x'}-2At'\frac{\partial}{\partial y'},\,\,\,\,\,\,\,\,\,\,\,\,\,\,\,\,\,\,\frac{\partial}{\partial y}\equiv\frac{\partial}{\partial y'},
\notag\\
\frac{\partial}{\partial t}\equiv\frac{\partial}{\partial t'}-2Ax'\frac{\partial}{\partial y'},\,\,\,\,\,\,\,\,\,\,\,\,\,\,\,\,\,\,\frac{\partial}{\partial z}\equiv\frac{\partial}{\partial z'}.
\end{align}
In this sheared system of coordinates, all disc quantities as perturbed as  $\chi=\chi_{\rm 0}+\delta \chi$, where $\delta \chi$ is assumed to be small in comparison to the initial state. The disturbances, denoted by primes,   are proportional to $\exp[i(k_{x}x'+k_{y}y')]$, where $k_{x}$ and $k_{y}$ are the radial and azimuthal wavenumbers respectively. Thus, the equations (8)-(14) become
\begin{equation}
\frac{\partial(\delta\Sigma_{\rm g})}{\partial t'}+\Sigma_{\rm 0g}(ik_{x}-2ik_{y}At')\delta v_{x{\rm g}}+ik_{y}\Sigma_{\rm 0g}(\delta v_{y{\rm g}})=0,
\end{equation}

\begin{align}
\frac{\partial(\delta v_{x{\rm g}})}{\partial t'}-2\Omega(\delta v_{y{\rm g}})=(ik_{x}-2ik_{y}At')[-(\delta\psi_{\rm g}+\delta\psi_{\rm d})
\notag\\
-\frac{c_{\rm s}^2}{\Sigma_{\rm 0g}}(\delta\Sigma_{\rm g})]+\frac{\epsilon(\delta v_{x{\rm d}}-\delta v_{x{\rm g}})}{t_{\rm stop}}-\frac{4}{3}\nu(k_{x}-2k_{y} At')^2(\delta v_{x{\rm g}})
\notag\\
-\nu k_{y}^2(\delta v_{x{\rm g}})+\frac{2iA k_{ y}\nu}{\Sigma_{\rm 0g}}(\delta \Sigma_{\rm g})-\nu k_{y}(k_{x}-2k_{y} At')(\delta v_{y{\rm g}}),
\end{align}

\begin{align}
\frac{\partial(\delta v_{y{\rm g}})}{\partial t\prime}+2B(\delta v_{x{\rm g}})=ik_{y}[-(\delta\psi_{\rm g}+\delta\psi_{\rm d})-\frac{c_{\rm s}^2}{\Sigma_{\rm 0g}}(\delta\Sigma_{\rm g})]
\notag\\
+\frac{\epsilon(\delta v_{y{\rm d}}-\delta v_{y{\rm g}})}{t_{\rm stop}}-\frac{4}{3}\nu k_{y}^2(\delta v_{y{\rm g}})-\nu(k_{x}-2k_{y}At')^2(\delta v_{y{\rm g}})
\notag\\
+\frac{2A\nu}{\Sigma_{\rm 0g}}(ik_{x}-2ik_{y}At')(\delta \Sigma_{\rm g})-\nu k_{y} (k_{x}-2k_{y}At') (\delta v_{x{\rm g}}),
\end{align}

\begin{align}
\frac{\partial(\delta\Sigma_{\rm d})}{\partial t'}+\Sigma_{\rm 0d}(ik_{x}-2ik_{y}At')\delta v_{x{\rm d}}+ik_{y}\Sigma_{\rm 0d}(\delta v_{y{\rm d}})
\notag\\
=D[-(k_{x}-2k_{y}At')^2-k_{y}^2]\delta\Sigma_{\rm d},
\end{align}

\begin{align}
\frac{\partial(\delta v_{x{\rm d}})}{\partial t'}-2\Omega(\delta v_{y{\rm d}})=(ik_{x}-2ik_{y}At')[-(\delta\psi_{\rm g}+\delta\psi_{\rm d})
\notag\\
-\frac{c_{\rm d}^2}{\Sigma_{\rm 0d}}(\delta\Sigma_{\rm d})]+\frac{(\delta v_{x{\rm g}}-\delta v_{x{\rm d}})}{t_{\rm stop}},
\end{align}

\begin{align}
\frac{\partial(\delta v_{y{\rm d}})}{\partial t'}+2B(\delta v_{x{\rm d}})=ik_{y}[-(\delta\psi_{\rm g}+\delta\psi_{\rm d})-\frac{c_{\rm d}^2}{\Sigma_{\rm 0d}}(\delta\Sigma_{\rm d})]
\notag\\
+\frac{(\delta v_{y{\rm g}}-\delta v_{y{\rm d}})}{t_{\rm stop}},
\end{align}

\begin{align}
[-(k_{x}-2Ak_{y}t')^2-k_{y}^2+\frac{\partial^2}{\partial z'^2}](\delta\psi_{g}+\delta\psi_{d})
\notag\\
=4\pi G(\delta\Sigma_{g}+\delta\Sigma_{d})\delta(z').
\end{align}

For perturbations with a non-zero $k_{y}$, it is more convenient to re-write equations (17)-(23) in terms of a    new dimensionless time variable, i.e.   $\tau=2At'-k_{ x}/k_{ y}$ \citep{gold}. Therefore, we obtain
\begin{equation}\label{eq:main-1}
\frac{\partial(\delta\Sigma_{\rm g})}{\partial\tau}+i\frac{k_{ y}}{2A}\Sigma_{\rm 0g}(\delta v_{y {\rm g}}-\tau \delta v_{x{\rm g}})=0,
\end{equation}

\begin{align}\label{eq:main-2}
\frac{\partial(\delta v_{x\rm g})}{\partial\tau}-\frac{\Omega}{A}(\delta v_{y\rm g})=-i\frac{k_{ y}}{2A}\tau[-(\delta\psi_{\rm g}+\delta\psi_{\rm d})-\frac{c_{\rm s}^2}{\Sigma_{\rm 0g}}(\delta\Sigma_{\rm g})]
\notag\\
+\frac{\epsilon(\delta v_{x\rm d}-\delta v_{x\rm g})}{2A t_{\rm stop}}-\frac{4}{3} \frac{k_{y}^2 \tau^2 \nu}{2A}  (\delta v_{x \rm g})-\frac{\nu k_{ y}^2}{2A}(\delta v_{x \rm g})
\notag\\
+\frac{i k_{y} \nu}{\Sigma_{\rm 0g}}(\delta \Sigma_{\rm g})+ \frac{k_{y}^2 \nu \tau}{2A} (\delta v_{y \rm g}),
\end{align}

\begin{align}\label{eq:main-3}
\frac{\partial(\delta v_{y\rm g})}{\partial\tau}+\frac{B}{A}(\delta v_{x\rm g})=i\frac{k_{ y}}{2A}[-(\delta\psi_{\rm g}+\delta\psi_{\rm d})-\frac{c_{\rm s}^2}{\Sigma_{\rm 0g}}(\delta\Sigma_{\rm g})]
\notag\\
+\frac{\epsilon(\delta v_{y\rm d}-\delta v_{y\rm g})}{2A t_{\rm stop}}-\frac{4}{3} \frac{k_{y}^2 \nu}{2A} (\delta v_{y \rm g})-\frac{k_{ y}^2 \nu  \tau^2}{2A}(\delta v_{y \rm g})
\notag\\
-\frac{i k_{y} \nu \tau}{\Sigma_{\rm 0g}}(\delta \Sigma_{\rm g})+ \frac{k_{y}^2 \nu \tau}{2A} (\delta v_{x \rm g}),
\end{align}

\begin{equation}\label{eq:main-4}
\frac{\partial(\delta\Sigma_{\rm d})}{\partial\tau}+i\frac{k_{ y}}{2A}\Sigma_{\rm 0d}(\delta v_{y\rm d}-\tau\delta v_{x\rm d})=-\frac{D k_{y}^2}{2A}(1+\tau^2)\delta\Sigma_{\rm d},
\end{equation}

\begin{align}\label{eq:main-5}
\frac{\partial(\delta v_{x\rm d})}{\partial\tau}-\frac{\Omega}{A}(\delta v_{y\rm d})=-i\frac{k_{y}}{2A}\tau[-(\delta\psi_{\rm g}+\delta\psi_{\rm d})-\frac{c_{\rm d}^2}{\Sigma_{\rm 0d}}(\delta\Sigma_{\rm d})]
\notag\\
+\frac{(\delta v_{x \rm g}-\delta v_{x \rm d})}{2A t_{\rm stop}},
\end{align}

\begin{align}\label{eq:main-6}
\frac{\partial(\delta v_{y\rm d})}{\partial\tau}+\frac{B}{A}(\delta v_{x\rm d})=i\frac{k_{y}}{2A}[-(\delta\psi_{\rm g}+\delta\psi_{\rm d})-\frac{c_{\rm d}^2}{\Sigma_{\rm 0d}}(\delta\Sigma_{\rm d})]
\notag\\
+\frac{(\delta v_{y\rm g}-\delta v_{y\rm d})}{2A t_{\rm stop}},
\end{align}

\begin{equation}\label{eq:pois}
[-k_{ y}^2(1+\tau^2)+\frac{\partial^2}{\partial z\prime^2}](\delta\psi_{\rm g}+\delta\psi_{\rm d})=4\pi G(\delta\Sigma_{\rm g}+\delta\Sigma_{\rm d})\delta(z').
\end{equation}
Upon solving the Poisson equation (\ref{eq:pois}) and using the approximation of the finite thickness of the disc \citep{Vandervoort1970,Shu1984}, we obtain the gravitational potential perturbation:
\begin{align}\label{eq:main-7}
(\delta\psi_{\rm g}+\delta\psi_{\rm d})=-\left (\frac{2\pi G}{k_{ y}(1+\tau^2)^\frac{1}{2}} \right )[\frac{\delta\Sigma_{\rm g}}{1+k_{ y}(1+\tau^2)^\frac{1}{2} H}
\notag\\
+\frac{\delta\Sigma_{\rm d}}{1+k_{ y}(1+\tau^2)^\frac{1}{2} H_{\rm d}}].
\end{align}
where $H=c_{\rm s}/\Omega $ and $H_{\rm d}=\sqrt{(\alpha /{\rm St})H}$ are the gas disc scale height and the dust scale height respectively.

Equations (\ref{eq:main-1})-(\ref{eq:main-6}) and (\ref{eq:main-7}) constitute main equations of the model to be solved subject to appropriate initial conditions. In the absence of drag force, mathematical forms of these equations are  similar to \cite{Jog1992}  who studied GI in a two-component disc consisting of the gas and stars to mimic a galaxy. Our second component, however, is the dust fluid and its feedback on the gas is included via the drag force. 
Time evolution of the perturbations is studied by solving our main equations.  For numerical purposes, however, it is better to re-write equations (\ref{eq:main-1})-(\ref{eq:main-6}) and (\ref{eq:main-7}) in terms of dimensionless variables. In doing so, we introduce the following new variables: 
\begin{align}
Q_{\rm g}=\frac{\kappa c_{\rm s}}{\pi G\Sigma_{\rm 0g}},\,\,\,\,\,\,\,\,\, Q_{\rm d}=\frac{\kappa c_{\rm d}}{\pi G\Sigma_{\rm 0d}},\,\,\,\,\,\,\,\,\,\,\,\,\,\,\,\eta=\frac{2A}{\Omega_{\rm 0}}
\notag\\
\delta u_{x\rm g}=\frac{\delta v_{x\rm g}}{c_{\rm s}},\,\,\,\,\,\,\,\delta u_{y\rm g}=\frac{\delta v_{y\rm g}}{c_{\rm s}},\,\,\,\,\,\,\,\,\,\,\,\delta u_{x\rm d}=\frac{\delta v_{x\rm d}}{c_{\rm s}},\,\,\,\,\,\,\
\notag\\
\delta u_{y\rm d}=\frac{\delta v_{y\rm d}}{c_{\rm s}},\,\,\,\,\,\,\,\theta_{\rm g}=\frac{\delta\Sigma_{\rm g}}{\Sigma_{\rm 0g}},\,\,\,\,\,\,\,\,\theta_{\rm d}=\frac{\delta\Sigma_{\rm d}}{\Sigma_{\rm 0d}}
\notag\\
R^2=\frac{\kappa^2}{4A^2}=\frac{2(2-\eta)}{\eta^2},\,\,\,\,\,\,\,\,\,\,X=\frac{\lambda_{y}}{\lambda_{\rm crit}}.
\end{align}
Here, parameters $Q_{\rm g}$ and $Q_{\rm d}$ stand for the Toomre parameter for the gas and dust components respectively. The  epicyclic frequency is denoted by $\kappa$ where for a Keplerian disc it becomes $\kappa=\Omega$. Furthermore,  the shear parameter is $\eta={2A}/{\Omega_{\rm 0}}$. Ratios of the gas and   dust surface densities and their corresponding initial values are shown by $\theta_{\rm g}$ and $\theta_{\rm d}$ respectively. We also introduce a critical wavelength, i.e. $\lambda_{\rm crit}=4\pi^2 G\Sigma_{\rm 0g}(1+\epsilon)/\kappa^2$, and the perturbation wavelength is written in terms of this critical wavelength. 

Using introduced dimensionless variables, equations (\ref{eq:main-1})-(\ref{eq:main-6}) and (\ref{eq:main-7}) are reduced to the following set of ordinary differential equations:

\begin{figure*}
\centering
\includegraphics[scale=1.0]{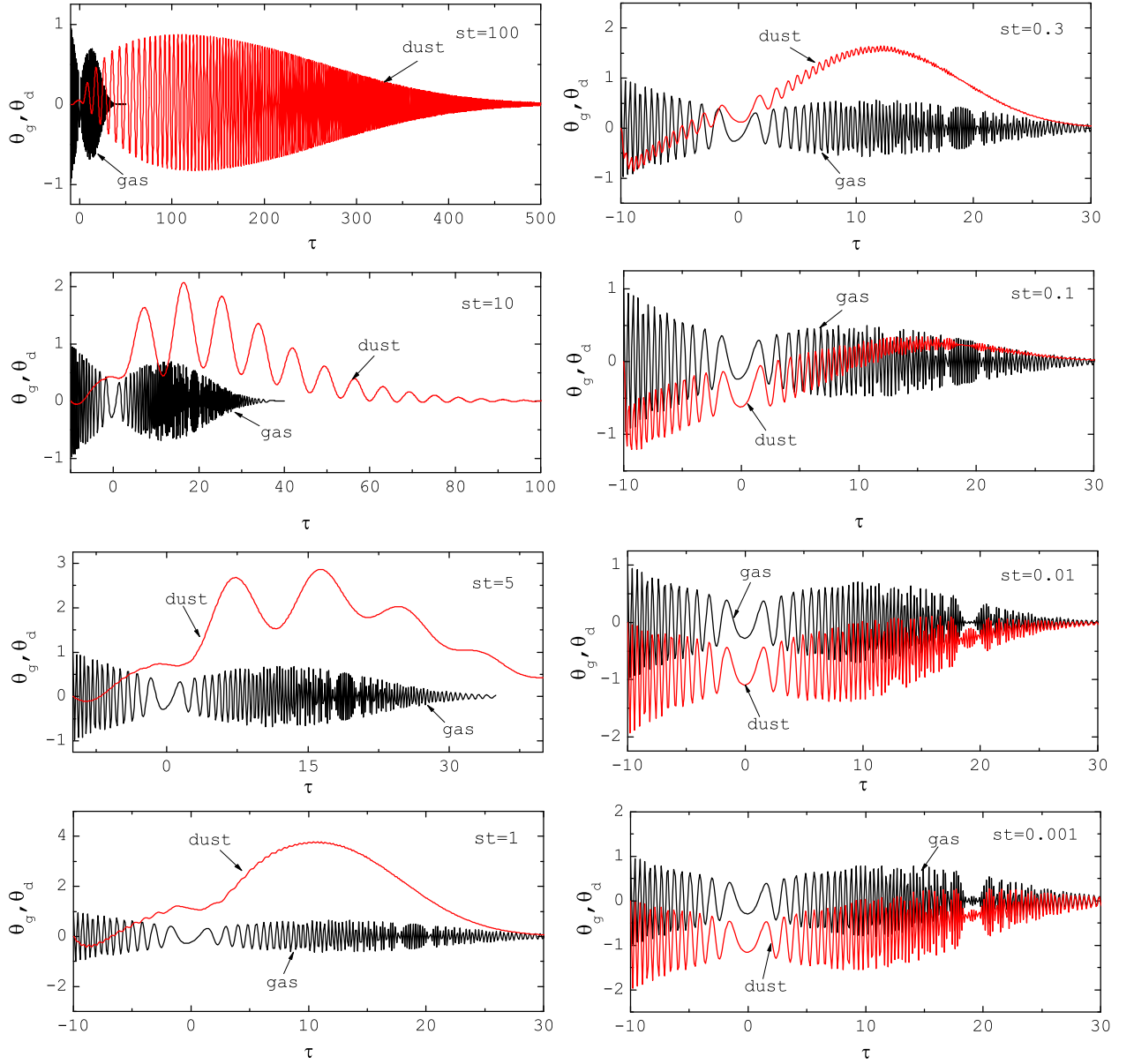}
\caption{The ratio of the  surface density perturbation to the unperturbed surface density for the gas (black curve) and dust (red curve) components, i.e. $\theta_{\rm g}$ and $\theta_{\rm d}$ versus the dimensionless time parameter $\tau$ subject to the initial conditions (\ref{eq:IC}) with  $\tau_{\rm ini}=-10$. Other model parameters are $Q_{\rm g}=15$, $Q_{\rm d}=11$, $\epsilon=0.01$, $X=3$, $\eta=1.5$ and $\alpha=10^{-4}$. }
\label{Fig1}
\end{figure*}

\begin{figure*}
\centering
\includegraphics[scale=0.6]{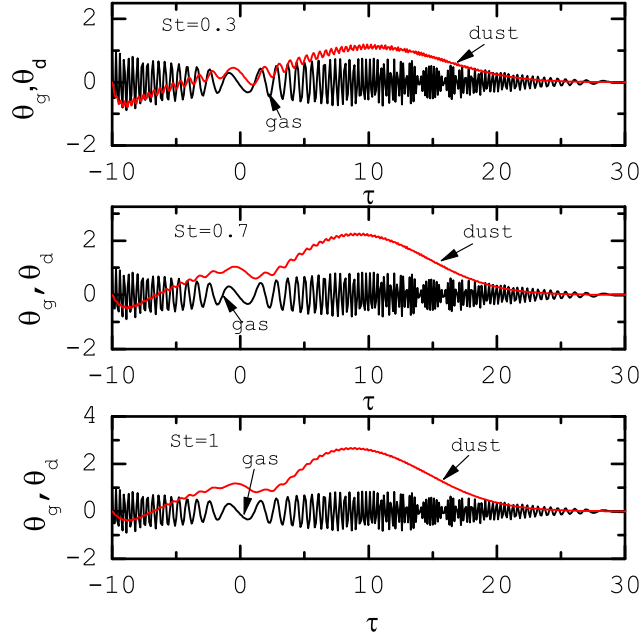}
\caption{Evolution of the gas surface density perturbation (black curve) and the dust surface density perturbation (red curve) in MMSN  at the radial distance 100 au and for different Stokes numbers, as labeled. The input parameters are $Q_{\rm g}=17.7$, $\epsilon=0.01$, $X=2.78$, $\eta=1.5$ and $\alpha=10^{-4}$. Also, the dust Toomre parameter is $Q_{\rm d}=18.24$, $18.43$ and $17.89$ corresponding to the Stokes numbers  $1$, $0.7$ and $0.3$, respectively. }
\label{Fig2}
\end{figure*}

\begin{align}\label{eq:Mnu-a}
-\frac{d \theta_{\rm g}}{d\tau}[\frac{\eta-2}{\eta}+\frac{1}{6}\frac{\alpha Q_{\rm g}^2 \eta R^2}{ X^2 (1+\epsilon)^2}\frac{\tau (1-\tau^2)}{1+\tau^2} ]
\notag\\
-\theta_{\rm g}(\frac{\alpha Q_{\rm g}^2 \eta R^2}{4 X^2 (1+\epsilon)^2})(1-\tau^2)-\frac{d W_{\rm g}}{d\tau}(\frac{Q_{\rm g}R}{2X(1+\epsilon)})
\notag\\
-W_{\rm g}[\frac{R \epsilon Q_{\rm g}}{2\eta \rm St X (1+\epsilon)}+\frac{\alpha Q_{\rm g}^3 \eta R^3}{8 X^3 (1+\epsilon)^3}(\frac{(1-\tau^2)^2}{1+\tau^2}+\frac{8 \tau^2}{3(1+\tau^2)})]
\notag\\
+W_{\rm d}(\frac{R \epsilon Q_{\rm g}}{2\eta \rm St X (1+\epsilon)}) =0
\end{align}

\begin{align}
-\frac{d^2 \theta_{\rm g}}{d \tau^2}+\frac{d \theta_{\rm g}}{d \tau}[\frac{2\tau}{1+\tau^2}-\frac{\epsilon}{\eta {\rm St}}-\frac{\alpha Q_{\rm g}^2 \eta R^2}{ X^2 (1+\epsilon)^2}(\frac{1+\tau^4}{3(1+\tau^2)}
\notag\\
+\frac{\tau^2}{1+\tau^2})]+\theta_{\rm g}[-\frac{\alpha Q_{\rm g}^2 \eta R^2}{2 X^2 (1+\epsilon)^2}\tau-\frac{Q_{\rm g}^2 R^2}{4 X^2 (1+\epsilon)^2}(1+\tau^2)
\notag\\
+\frac{R^2}{X(1+\epsilon)}(\frac{(1+\tau^2)^{1/2}}{1+\frac{Q_{\rm g}}{2X(1+\epsilon)}(1+\tau^2)^{1/2}})]+\frac{d \theta_{\rm d}}{d \tau}(\frac{\epsilon}{\eta {\rm St}})
\notag\\
+\theta_{\rm d}[\frac{R^2\epsilon}{X(1+\epsilon)}(\frac{(1+\tau^2)^{1/2}}{1+\sqrt{\frac{\alpha}{\rm St}}\frac{Q_{\rm g}}{2X(1+\epsilon)}(1+\tau^2)^{1/2}})
\notag\\
+\frac{\epsilon\xi Q_{\rm g}^2 R^2}{4X^2 {\rm St}(1+\epsilon)^2}(1+\tau^2)]+W_{\rm g}[\frac{Q_{\rm g} R}{X(1+\epsilon)}(\frac{1}{1+\tau^2})
\notag\\
-\frac{R Q_{\rm g}}{X \eta (1+\epsilon)}-\frac{1}{12}\frac{\alpha Q_{\rm g}^3 \eta R^3}{X^3 (1+\epsilon)^3}\frac{\tau(1-\tau^2)}{1+\tau^2}]=0
\end{align}

\begin{align}
-\frac{d W_{\rm d}}{d\tau}(\frac{Q_{\rm g} R}{2X(1+\epsilon)})-\frac{d \theta_{\rm d}}{d \tau}(\frac{\eta-2}{\eta})
\notag\\
-\theta_{\rm d}(\frac{\xi Q_{\rm g}^2 R^2 (\eta-2)}{4 X^2(1+\epsilon)^2})(1+\tau^2)+\frac{R Q_{\rm g}}{2\eta {\rm St X} (1+\epsilon)}(W_{\rm g}-W_{\rm d})=0
\end{align}

\begin{align}\label{eq:Mnu-b}
-\frac{d^2\theta_{\rm d}}{d\tau^2}+\frac{d \theta_{\rm d}}{d\tau}[\frac{2\tau}{1+\tau^2}-\frac{1}{\eta {\rm St}}-\frac{\xi \eta Q_{\rm g}^2 R^2 }{4X^2(1+\epsilon)^2}(1+\tau^2)]
\notag\\
+\frac{d \theta_{\rm g}}{d\tau}(\frac{1}{\eta {\rm St}})+\theta_{\rm d}[\frac{R^2\epsilon}{X(1+\epsilon)}(\frac{(1+\tau^2)^{1/2}}{1+\sqrt{\frac{\alpha}{\rm St}}\frac{Q_{\rm g}}{2X(1+\epsilon)}(1+\tau^2)^{1/2}})
\notag\\
-\frac{\xi Q_{\rm g}^2 R^2}{4X^2 {\rm St} (1+\epsilon)^2}(1+\tau^2)-(\frac{\epsilon Q_{\rm d} R}{2 X (1+\epsilon)})^2 (1+\tau^2)]
\notag\\
+\theta_{\rm g}[\frac{R^2}{X(1+\epsilon)}(\frac{(1+\tau^2)^{1/2}}{1+\frac{Q_{\rm g}}{2X(1+\epsilon)}(1+\tau^2)^{1/2}})]
\notag\\
+W_{\rm d}[\frac{Q_{\rm g} R}{X (1+\epsilon)}(\frac{1}{1+\tau^2})-\frac{R Q_{\rm g}}{X \eta (1+\epsilon)}]=0
\end{align}
where $W_{\rm g}=i[\delta u_{x\rm g}+\tau(\delta u_{y\rm g}) ]$ and $W_{\rm d}=i[\delta u_{x\rm d}+\tau(\delta u_{y\rm d})]$. Equations (\ref{eq:Mnu-a})-(\ref{eq:Mnu-b}) are solved using Runge-Kutta method to determine evolution of the perturbations with time. In the next section, we present our solutions.

\section{Numerical solutions}

Although a wide range of the initial conditions can be implemented, we consider the following simple initial conditions. At the initial time $\tau_{\rm ini}$, the relevant quantities are 
\begin{equation}\label{eq:IC}
(\theta_{\rm g}, d \theta_{\rm g}/d \tau, \theta_{\rm d}, d \theta_{\rm d}/d \tau, W_{\rm g}, W_{\rm d})=(1,0,0,0,0,0).
\end{equation}
When a perturbation  oscillates with an amplitude less than unity, we consider it as a stable configuration. Note that stable axisymmetric  perturbations also exhibit oscillatory behavior with time. The non-axisymmetric perturbations, however, are unstable once their amplitudes display significant growth only for a limited time. Unstable axisymmetric perturbations grow with an exponential profile, but growth of the nonaxisymmetric perturbations  is not exponential   and the concept of the instability corresponds to rapid transient amplification of the  perturbations during a limited time. These transient amplifications have already been found in the linear stability analysis of the gaseous self-gravitating discs \citep{Mamats2013},  discs with gas and stars \citep{Jog1992}  and even nonaxisymmetric MRI \citep{Balbus1992}. It is therefore quite normal that a mixture of gas and dust undergoes transient amplifications subject to the nonaxisymmetric perturbations. The fate of these transient patterns can be addressed in the non-linear regime, however, it is important to specify range of the model parameters for which the system becomes linearly unstable. We thereby consider initial states which are stable subject to the axisymmetric perturbations. But are they remain stable subject to the nonaxisymmetric perturbations? This is an important question that motivated us to perform stability analysis with a broad range of the model parameters.

We first investigate the evolution of the perturbations in Figure \ref{Fig1} for a fiducial set of the model parameters. We consider an initial configuration with the gas and dust large Toomre parameters to ensure  stability of the system against to the axisymmetric perturbations. The model parameters, therefore, are assumed  $Q_{\rm g}=15$, $Q_{\rm d}=11$, $\epsilon=0.01$, $ X=3$, $\eta=1.5$ and $\alpha=10^{-4}$. The initial dimensionless time is $\tau_{\rm ini}=-10$. Note that we also verified that for other values of $\tau_{\rm ini}$ the results are qualitatively similar. We, however, note that the initial time $t=0$ corresponds to $\tau_{\rm ini} = -k_{x}/k_{y}$ that can be rewritten as $k_x = -\tau_{\rm ini} k_y$. Since the azimuthal wavenumber $k_y$ is a given model parameter in terms of the dimensionless parameter $X$, we can infer that perturbations with a larger $|\tau_{\rm ini}|$ have a larger radial wavenumber.  Each panel of Figure \ref{Fig1} shows the evolution of the gas and dust perturbations, i.e. $\theta_{\rm g}$ and $\theta_{\rm d}$, as a function of $\tau$ for a given Stokes number, as labeled. Obviously, a given Stokes number is equivalent to a certain dust size. Black and red solid curves correspond to the gas and dust components respectively.

Figure \ref{Fig1} shows that the gas component is always stable irrespective of the Stokes number. Since the gas Toomre parameter is larger than its critical value, the gas component is stable subject to the perturbations and dust dynamics is unable to  change this trend due to a small dust-to-gas density ratio. The response of the dust component, however, strongly depends on the adopted Stokes number that controls the magnitude of the drag force. For a large Stokes number, the dust component is also stable because the drag force is too weak to affect dust dynamics. But when the Stokes number is less than about 10, the dust component tends to be unstable subject to the nonaxisymmetric perturbations whereas the gas is still stable. We also find that the dust component is unstable for ${\rm St}=0.3$. If the  Stokes number is adopted less than about 0.3, not only the dust component becomes stable but also its evolution is  similar to the gas evolutionary trend. This behavior is understood in terms of the strong dust and gas coupling for the small Stokes numbers. Under this condition, the stability of the gas component is dictated to the dust component due to their strong coupling. Thus, for an intermediate range of the Stokes number $0.3\lesssim {\rm St} \lesssim 10$, while the gas component remains stable, the dust component undergoes transient growing patterns during a time interval almost independent of the gas component. But either the dust component undergoes a growing phase or just display oscillatory behaviour, the amplitudes of the perturbations decay at larger times.

In a realistic case, however, our model parameters can not be adopted independently as we did in Figure \ref{Fig1}. These parameters depend on the gaseous disc model and the initial distribution of the dust particles. In agreement with most previous studies in this context, we consider the minimum mass solar nebula \citep[MMSN;][]{hayashi81} to represent our gaseous disc model. However, the entire structure of a PPD is unlikely to be described using the MMSN model  and there are also alternative disc profiles \citep[][]{Nixon2018}. Our stability analysis, therefore, is done in an MMSN model at a given radial distance. In this model, the surface density and sound speed are given as power-law functions of the radial distance \citep{hayashi81}:
\begin{equation}
\Sigma(r)=1.7 \times 10^{3} \left ( \frac{r}{1 \rm au} \right )^{-\frac{3}{2}}     \,\,\,\,\,{\rm g \hspace{1mm} cm^{-2}},
\end{equation}
\begin{equation}
c_{\rm s}(r)=1.0 \times 10^{5} \left ( \frac{r}{1 \rm au} \right )^{-\frac{1}{4}} \,\,\,\,\,{\rm cm \hspace{1mm}s}^{-1} .
\end{equation}
In MMSN model with a solar mass host star, the gas Toomre parameter can therefore be expressed by
\begin{equation}
Q_{\rm g}=56 \left ( \frac{r}{1 \rm au} \right )^{-\frac{1}{4}}.
\end{equation}
Furthermore, the Toomre parameter associated to the dust component becomes
\begin{equation}
Q_{\rm d} = \frac{56}{\epsilon} \alpha^{\frac{1}{2}} \left ( \frac{r}{1 \rm au} \right )^{-\frac{1}{4}} \left [\frac{1+2 (\rm St)^2 +5/4 (\rm St)^3}{(1+(\rm St)^2)^2} \right ]^{\frac{1}{2}}
\end{equation}
The Stokes number therefore at the disc midplane becomes
\begin{equation}
{\rm St}=1.8 \times 10^{-7} \left (\frac{a}{1 {\rm \mu m}} \right ) \left ( \frac{r}{1 {\rm au}} \right )^{\frac{3}{2}}.
\end{equation}
We can use the above relations for determining model parameters self-consistently in the MMSN model at a given radial location. In Table \ref{tab1}, we list our model parameters for the MMSN model at the radial distance 100 au. 

\begin{table}

\begin{center}

\begin{tabular}{c | c | c | c | c | c  }

\hline

$\epsilon$&      $\alpha$&	$Q_{\rm g}$&	$Q_{\rm d}$&	  $\rm St$&	$a(\mu m)$ \\ \hline

$0.01$&	 $10^{-4}$& $17.7$& $6.66$& $10$& $55555$\\

$0.01$&	$10^{-4}$& $17.7$& $18.24$& $1$& $5555$\\

$0.01$& $10^{-4}$& $17.7$& $17.71$& $0.1$& $555$\\ 

$0.01$& $10^{-4}$& $17.7$& $17.70$& $0.01$& $55$\\

$0.01$& $10^{-4}$& $17.7$& $17.70$& $0,001$& $5.5$\\

$0.01$& $10^{-4}$& $17.7$& $17.70$& $0.0001$& $0.5$\\

\end{tabular}

\end{center}

\caption{Our model  parameters in the MMSN model with a solar mass host star at the  radial distance 100 au.}

\end{table}\label{tab1}

\begin{figure*}
\centering
\includegraphics[scale=0.5]{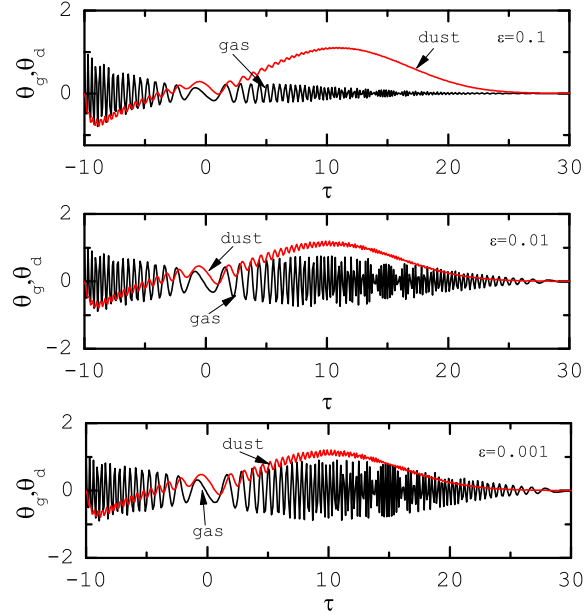}
\caption{The ratio of the perturbation surface density to the unperturbed surface density for the gas (black curve) and dust (red curve) components, i.e. $\theta_{\rm g}$ and $\theta_{\rm d}$ versus the dimensionless time parameter $\tau$ for different values of metalicity   in MMSN model at the radial distance 100 au. Evolution of the perturbations is calculated subject to the initial condition (\ref{eq:IC}) with $\tau_{\rm ini}=-10$.  Other model parameters are ${\rm St}=0.3$, $Q_{\rm g}=17.7$, $\eta=1.5$ and $\alpha=10^{-4}$. For $\epsilon = 0.1$, $0.01$ and $0.001$, we have $Q_{\rm d}=1.78, $ 17.89 and 212.6 and the corresponding azimuthal dimensionless parameter becomes $X=2.5$,  $2.78$ and $178.9$, respectively.  }
\label{Fig3}
\end{figure*}

\begin{figure}
\centering
\includegraphics[scale=0.6]{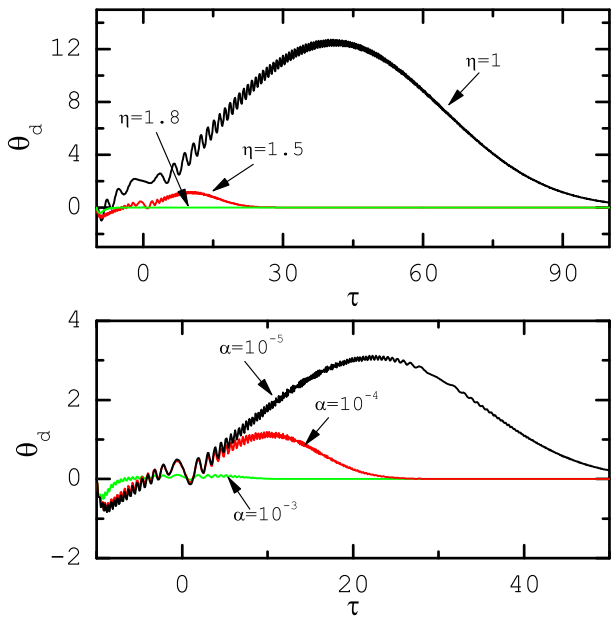}
\caption{The ratio of the dust surface density perturbation to the unperturbed surface density, i.e. $\theta_{\rm d}$, versus the dimensionless time parameter $\tau$ in MMSN model. Top panel is plotted for different values of the shear parameter $\eta$, whereas the bottom panel corresponds to different values of $\alpha$.  All input parameters are calculated at the radial distance 100 au. In both panels, we set ${\rm St}=0.3$, $\epsilon=0.01$ and $\tau_{\rm ini}=-10$. In the top panel, we have $\alpha=10^{-4}$ and each curve is labeled with the adopted shear parameter, i.e. $\eta=1$, 1.5 and 1.8. Corresponding to theses values, therefore, we have $Q_{\rm g}=249.9$, 17.7 and 2.8, $Q_{\rm d}=252.6$, 17.89 and 2.83 and $X=557.3$, 2.78 and 0.07, respectively.  In the bottom panel, we have   $Q_{\rm g}=17.7$, $X=2.78$ and $\eta=1.5$ and Toomre parameter of the dust component for $\alpha=10^{-3}$, $10^{-4}$ and $10^{-5}$ becomes $Q_{\rm d}=56.57$, $17.89$ and $5.65$ respectively. } 
\label{Fig4}
\end{figure}

\begin{figure}
\centering
\includegraphics[scale=0.6]{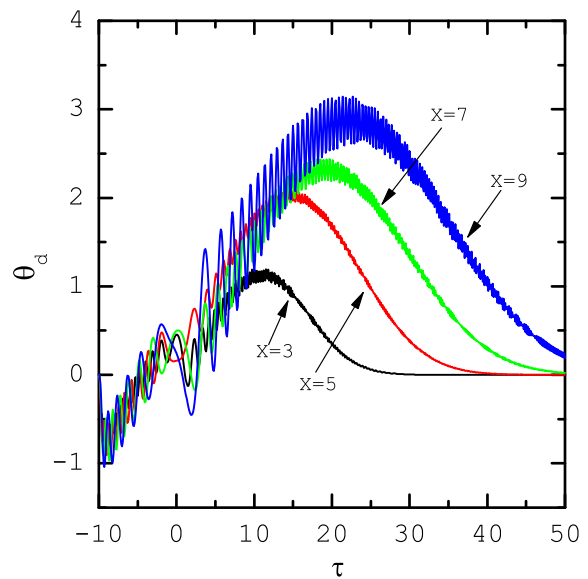}
\caption{The ratio of the dust surface density perturbation to the unperturbed surface density, i.e.  $\theta_{\rm d}$, versus the dimensionless time parameter $\tau$ in the MMSN model and for different values of $X$, as labeled. Rest of the model parameters are ${\rm St}=0.3$, $Q_{\rm g}=17.7$, $Q_{\rm d}=17.89$, $\epsilon=0.01$, $\eta=1.5$ and $\alpha=10^{-4}$. }
\label{Fig5}
\end{figure}

\begin{figure}
\centering
\includegraphics[scale=0.5]{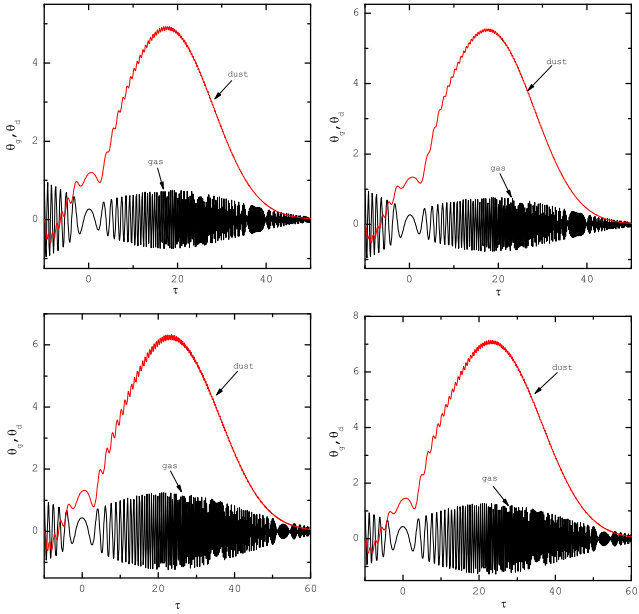}
\caption{The ratio of the surface density perturbation to the unperturbed surface density for the gas and dust components, i.e. $\theta_{\rm g}$ and $\theta_{\rm d}$ versus the dimensionless time parameter $\tau$ in MMSN model at the radial distance 100 au. As before, the initial condition (\ref{eq:IC}) is implemented  with  $\tau_{\rm ini}=-10$. In all panels, we have $X=2.78$, $\eta=1.5$, $\epsilon=0.01$  and $\alpha=10^{-4}$. But other model parameters are  ${\rm St}=0.7$, $Q_{\rm g}=7$ and $Q_{\rm d}=7.29$  (top-left), and ${\rm St}=0.9$, $Q_{\rm g}=7$ and $Q_{\rm d}=7.26$ (top-right),  ${\rm St}=0.7$, $Q_{\rm g}=5$ and  $Q_{\rm d}=5.2$ (bottom-left),  ${\rm St}=0.9$, $Q_{\rm g}=5$ and $Q_{\rm d}=5.2$  (bottom-right).}
\label{Fig6}
\end{figure}

\begin{figure}
\centering
\includegraphics[scale=0.5]{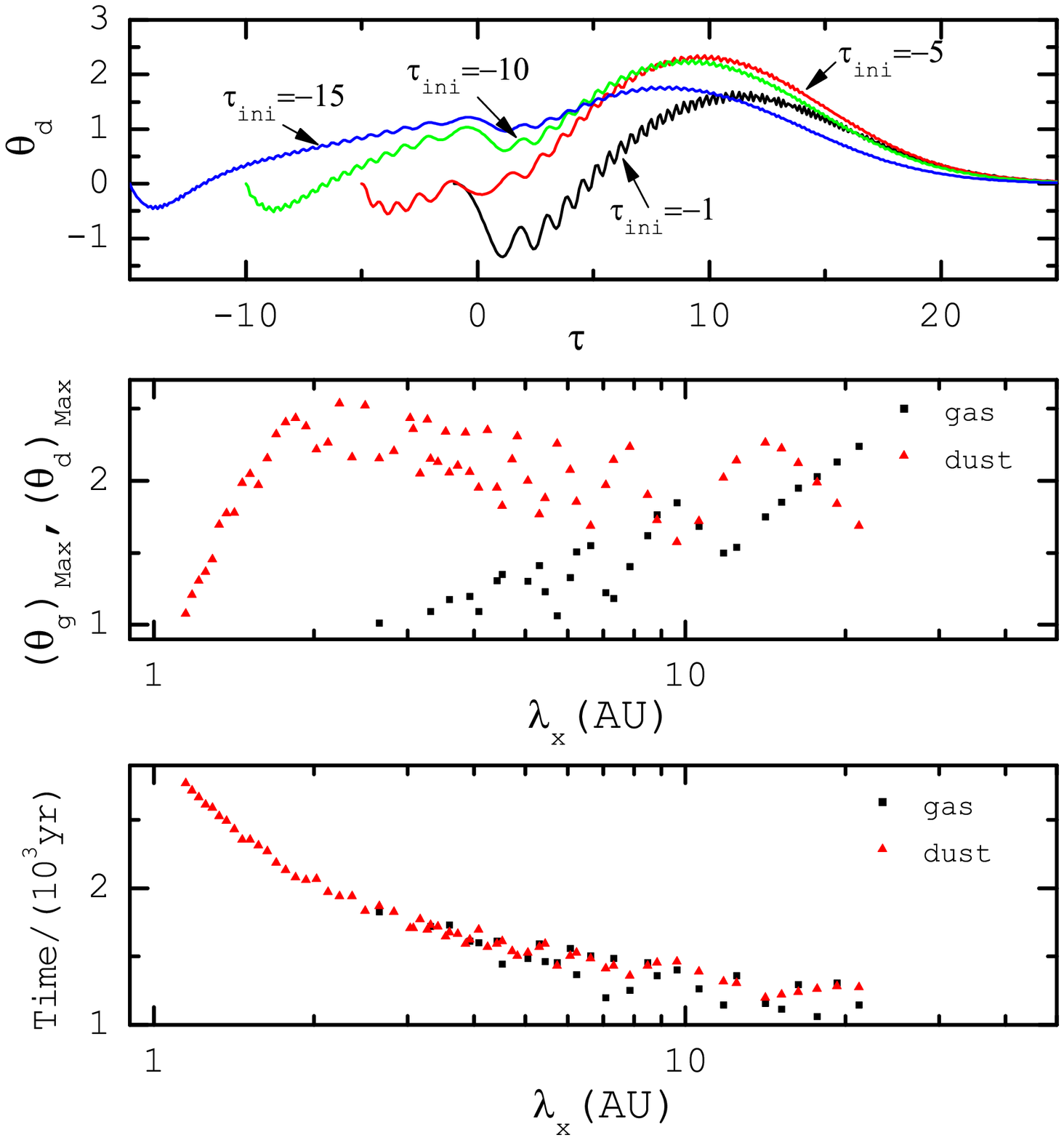}
\caption{Top panel shows evolution of dust component perturbations with different initial dimensionless time $\tau_{\rm ini}$, as labeled. The other model parameters are ${\rm St}=0.7$, $\epsilon =0.01$, $Q_{\rm g}=17.7$, $Q_{\rm d}=18.43$, $X=2.78$, $\eta=1.5$ and $\alpha =10^{-4}$. Middle panel displays the maximum growth amplitude for the gas (black square symbol) and the dust components (red triangle symbol) as a function of the radial perturbation wavelength $\lambda_{\rm x}$ for the model parameters same as in figure \ref{Fig2} with ${\rm St}=0.7$. In the bottom panel, the associated growth time (i.e., the time when the amplitude becomes maximum) is shown as a function of the radial perturbation wavelength.}
\label{Fig7}
\end{figure}

Figure \ref{Fig2} displays the evolution of the gas surface density perturbation (black curve)  and the dust surface density (red curve) in MMSN model with $M=1$ M$_{\odot}$ at the radial distance 100 au and for different Stokes numbers, as labeled. The gas Toomre parameter is   $Q_{\rm g}=17.7$, whereas the Toomre parameter for the dust component becomes  $Q_{\rm d} = 18.24$, 18.43 and 17.89 corresponding to the Stokes numbers 1, 0.7 and 0.3, respectively. Note that for these Stokes numbers, dust particles are millimeter-sized in a range between 1.6 mm and 5.5 mm.  We have chosen other model parameters as  $\epsilon=0.01$, $ X=2.78$, $\eta=1.5$ and $\alpha=10^{-4}$. We verified that both the gas and dust fluids are stable subject to the axisymmetric disturbances. But the response of the system to the nonaxisymmetric perturbations is quite different as we see in Figure \ref{Fig2}. While the gas component remains stable subject to  these perturbations irrespective of the adopted Stokes number, the dust component displays noticeable growth and the amplification factor reduces  with decreasing the Stokes number from 1 to 0.3.   

The dust abundance is a key parameter and its role is explore in Figure \ref{Fig3} for different values of $\epsilon$, as labeled. As in the previous figure, the model parameters correspond to MMSN model at the radial distance 100 au. We set ${\rm St}=0.3$, $Q_{\rm g}=17.7$, $\eta=1.5$ and $\alpha=10^{-4}$. For $\epsilon = 0.1$, $0.01$ and $0.001$, we have $Q_{\rm d}=1.78$, 17.89 and 178.9 and the corresponding azimuthal dimensionless parameter becomes $X=2.5$,   $2.78$ and $2.8$, respectively.  As before, the integration is started at $\tau_{\rm ini}=-10$ and the initial condition (\ref{eq:IC}) is implemented to determine evolution of the perturbations. The gas component exhibits an oscillatory behavior, however, its amplitude significantly decreases with increasing dust abundance. But the dust component undergoes a slightly higher growth rate when its abundance is increased.

In Figure \ref{Fig4}, we investigate the role of the parameters $\eta$ and $\alpha$ in the evolution of the disturbances. Model parameters correspond to MMSN model at the radial distance 100 au and the runs are started at $\tau_{\rm ini} =-10$. In both panels, we adopt ${\rm St}=0.3$ and $\epsilon=0.01$. In the top panel, we have $\alpha=10^{-4}$ and different values of the shear parameter are considered, as labeled. For the adopted values of the shear parameter, i.e. $\eta=1$, 1.5 and 1.8, the associated gas and dust Toomre parameter become  $Q_{\rm g}=249.9$, 17.7 and 2.8, and, $Q_{\rm d}=252.6$, 17.89 and 2.83, and, we also have $X=557.3$, 2.78 and 0.07, respectively. In the bottom panel, we assume that  $Q_{\rm g}=17.7$, $X=2.78$ and $\eta=1.5$ and the dust Toomre parameter  for $\alpha=10^{-3}$, $10^{-4}$ and $10^{-5}$ becomes $Q_{\rm d}=56.57$, $17.89$ and $5.65$, respectively. Note that both the gas and dust components are stable subject to the axisymmetric disturbances. In the top panel of Figure \ref{Fig4}, role of $\eta$ that quantifies the differential rotation rate    is shown for different values of this parameter. All model parameters are similar to the bottom panel and the viscosity coefficient is  $\alpha=10^{-4}$. As the parameter $\eta$ decreases, the dust component tends to be more unstable. The bottom panel of Figure \ref{Fig4} shows that fate of the dust component against to the nonaxisymmetric perturbations strongly depends on the chosen viscosity coefficient. While for $\alpha=10^{-3}$, the dust fluid remains stable, the amplitude of the perturbations gradually increases as the viscosity coefficient tends to the smaller values. 

In our stability analysis, we introduced the dimensionless parameter $X$ that measures azimuthal wavelength of the perturbations in terms of a critical wavelength $\lambda_{\rm crit}$. So far this parameter has been included as a fixed value. In Figure \ref{Fig5}, we explore role of this parameter in evolution of the dust component in MMSN model at the radial distance 100 au for these model parameters: $Q_{\rm g}=17.7$, $Q_{\rm d}=21.26$, ${\rm St}=0.3$,  $\epsilon=0.01$, $\eta=1.5$, $\alpha=10^{-4}$. As before, the starting time is $\tau_{\rm ini}=-10$. Each curve is marked with the corresponding value of $X$. Comparison of these curves show that  the amplitude of the oscillations increases with increasing $X$.

In Figure \ref{Fig6}, we  consider typical cases that satisfy conditions of the axisymmetric stability \citep{Taka} but finite amplifications are found subject to the nonaxisymmetric perturbations. We set $X=2.78$, $\eta=1.5$, $\epsilon=0.01$ and $\alpha=10^{-4}$. In the top left hand panel of Figure \ref{Fig6}, variations in $\theta_{\rm g}$ and $\theta_{\rm d}$ with $\tau$ are shown for ${\rm St}=0.7$, $Q_{\rm g}=7$ and $Q_{\rm d}=7.29$. The dust component is weakly  coupled to the gas component and the corresponding perturbations exhibit a significant growth during a finite time period, whereas the gas component remains stable. In the top right hand panel, the model parameters are ${\rm St}=0.9$, $Q_{\rm g}=7$ and $Q_{\rm d}=7.26$. The dust component again displays a fairly strong amplification for this choice of the parameters. In the bottom panels, we explore stability of the cases with a smaller Toomre parameter.  In the bottom left hand panel, we set ${\rm St}=0.7$, $Q_{\rm g}=5$ and $Q_{\rm d}=5.2$, whereas the bottom right hand panel corresponds to a case with the same gas and dust Toomre parameter but with a slightly larger Stokes number, i.e. ${\rm St}=0.9$. Since the Toomre parameter associated with each component is smaller,   the amplitude of the perturbations are larger for both cases with Stokes numbers $\rm St=0.7$ and $\rm St=0.9$ . We find that amplitude of the nonaxisymmetric perturbations increases with time and then these oscillations are damping.

In all explored cases so far, we used a fixed initial dimensionless time $\tau_{\rm ini} =-10$ which then it corresponds to a given radial perturbation wavelength $\lambda_{x}$ if the azimuthal wavelength is treated as a given fixed value. This argument is based on an already introduced relation as follows  $k_x = -\tau_{\rm ini} k_y$. We now investigate evolution of the perturbations with different initial dimensionless time $\tau_{\rm ini}$. This analysis, thereby, corresponds to evolution of the perturbations with different radial perturbation wavelength for a given fixed azimuthal wavelength. In the top panel of Figure \ref{Fig7}, we exhibit evolution of the dust component for different values of the initial time $\tau_{\rm ini}$, as labeled. The other model parameters are ${\rm St}=0.7$, $\epsilon =0.01$, $Q_{\rm g}=17.7$, $Q_{\rm d}=18.43$, $X=2.78$, $\eta=1.5$ and $\alpha =10^{-4}$. Note that the implemented values $\tau_{\rm ini}=-1$, -5, -10 and -15 correspond to radial wavelength $\lambda_x =21.23$ au, 4.24 au, 2.12 au and 1.41 au, respectively. This plot shows that the amplitude of the perturbations gradually increases with time, however, its growth is suppressed after a certain period of time depending on the radial perturbation wavelength.
In the middle panel of Figure \ref{Fig7}, we consider cases with  different initial time $\tau_{\rm ini}$ in an  interval from $-1$ to $-20$ and the resulting maximum amplitude for the gas and dust components, i.e. $(\theta_{\rm g})_{\rm Max}$ and $(\theta_{\rm d})_{\rm Max}$ are shown as a function of the radial perturbation wavenumber $\lambda_{ x}$.  The adopted model parameters are $\rm St=0.7$, $Q_{\rm g}=17.7$, $Q_{\rm d}=18.43$, $\epsilon=0.01$, $X=2.78$, $\alpha=10^{-4}$ and for a Keplerian disc with $\eta=1.5$. We find that for $\lambda_{ x} \leq 10$ au the dust component generally undergoes a larger $(\theta_{\rm d})_{\rm Max}$ in comparison to the gas component. In other words, the dust component is more unstable in comparison to the gas component for the short radial perturbation wavelength. In the bottom panel, we display growth time (i.e., the time when the amplitude becomes maximum) as a function of the radial perturbation wavelength for the explored cases.  It shows that both dust and gas component evolves to their maximum amplitude during more or less the same time period. This time scale, however, is in an interval from 1000 yr (for long wavelengths) to 3000 yr (for short wavelengths).  We find that evolution of the nonaxisymmetric growth time scale is relatively fast in comparison to the axisymmeric growth time scale.

\section{Conclusions}
We investigated evolution of the imposed nonaxisymmetric perturbations in a PPD by treating the system as a mixture of the coupled gas and dust particles. While response of the system subject to the axisymmetric perturbations may become unstable with an exponential growth, we find that nonaxisymmetric perturbations evolve with an oscillatory amplitude. But a strong amplification is found corresponding to the cases where are stable against to the axisymmetric perturbations. We can now summarize our main results:

- Growth of the nonaxisymmetric perturbations is not significant when the Stokes number is very large or very small. But there is always an intermediate range for the Stokes number where a high amplification in dust against to the nonaxisymmetric perturbations is found even when the gas component remains stable. However, the upper and lower limits of the Stokes number corresponding to the instability depend on the other model parameter including dust and gas Toomre parameters.

- In the MMSN model with a solar mass host star, we  found that amplitude of the nonaxisymmetric perturbations at the radial distance 100 au increases when the Stokes number lies in a range between $10^{-4}$ and $10^{-2}$. The amplification factor, however, decreases with increasing the Stokes number.

- Perturbations with a larger azimuthal wavelength exhibit a relatively higher amplification. 

- Turbulent coefficient has a stabilizing role in promoting nonaxisymmetric SGI. In the MMSN model, for example, we showed that growth of the perturbation is completely suppressed when the coefficient is $\alpha=10^{-3}$, whereas for smaller values of this coefficient, the amplification factor increases.

The final outcome of the implemented nonaxisymmetric perturbations cannot be addressed using the present linear analysis. Our study demonstrates that nonaxisymmetric perturbations may lead to dust transient amplification for a wider range of the model parameters in comparison to the axisymmetric perturbations. In the context of the nonaxisymmetric gravitational instability in galaxies, many authors found  similar evolutionary trends in either purely gaseous discs or two-fluid (stars and gas) systems \citep[e.g.,][]{Julian1966,Jog1992,Fuchs2001,Mich2016,Ghosh18}. The imposed nonaxisymmetric perturbations get amplified during an initial time interval due to the mutual interplay between various physical agents including shear and self-gravity of the disc, but eventually this amplification is suppressed because of the disc shear. Our model also predicts a similar trend for the evolution of the dust component even when the gas component remains stable. Although our basic equations are similar to previous studies relevant to gravitational instability in galaxies, there is a significant difference because each component (gas or dust) is permitted to exchange momentum via the drag force. In other words, dust and gas components are gravitationally coupled and they are subject to the momentum exchange by the drag force. We then found that the drag force is able to promote temporal growth of the dust component even in the cases where both components are stable subject to the axisymmetric perturbations. Furthermore, amplitude amplification of the dust component may persists in configurations where the gaseous disc responds via oscillations with non growing amplitudes.     

The above mentioned findings thereby propose that nonaxisymmetric SGI will have a better chance to exist. If that is the case, does it mean the observed dust rings in PPDs are the final outcome of SGI? This important question can not be adequately addressed with a linear analysis because it is not clear if the resulting nonaxisymmetric perturbation evolves to spiral waves or collapsing fragments. But we found that growth time scale of the nonaxisymmetric SGI is very fast. It then implies that the resulting patterns are less axisymmetric and probably the observed dust rings are not caused by SGI. Further numerical simulations are needed to address this essential question.

\section*{Acknowledgements}
We are grateful to referee for a constructive report that helped us to improve the manuscript. MS is also grateful to Henrik Latter for his constructive comments.

\bibliographystyle{mnras}
\bibliography{reference} 

\begin{thebibliography}{}
\makeatletter
\relax
\def\mn@urlcharsother{\let\do\@makeother \do\$\do\&\do\#\do\^\do\_\do\%\do\~}
\def\mn@doi{\begingroup\mn@urlcharsother \@ifnextchar [ {\mn@doi@}
  {\mn@doi@[]}}
\def\mn@doi@[#1]#2{\def\@tempa{#1}\ifx\@tempa\@empty \href
  {http://dx.doi.org/#2} {doi:#2}\else \href {http://dx.doi.org/#2} {#1}\fi
  \endgroup}
\def\mn@eprint#1#2{\mn@eprint@#1:#2::\@nil}
\def\mn@eprint@arXiv#1{\href {http://arxiv.org/abs/#1} {{\tt arXiv:#1}}}
\def\mn@eprint@dblp#1{\href {http://dblp.uni-trier.de/rec/bibtex/#1.xml}
  {dblp:#1}}
\def\mn@eprint@#1:#2:#3:#4\@nil{\def\@tempa {#1}\def\@tempb {#2}\def\@tempc
  {#3}\ifx \@tempc \@empty \let \@tempc \@tempb \let \@tempb \@tempa \fi \ifx
  \@tempb \@empty \def\@tempb {arXiv}\fi \@ifundefined
  {mn@eprint@\@tempb}{\@tempb:\@tempc}{\expandafter \expandafter \csname
  mn@eprint@\@tempb\endcsname \expandafter{\@tempc}}}

\bibitem[\protect\citeauthoryear{{Adams}, {Ruden}  \& {Shu}}{{Adams}
  et~al.}{1989}]{Adams89}
{Adams} F.~C.,  {Ruden} S.~P.,   {Shu} F.~H.,  1989, \mn@doi [\apj]
  {10.1086/168187}, \href {http://adsabs.harvard.edu/abs/1989ApJ...347..959A}
  {347, 959}

\bibitem[\protect\citeauthoryear{{Akimkin}}{{Akimkin}}{2015}]{Akimkin2015}
{Akimkin} V.~V.,  2015, \mn@doi [Astronomy Reports]
  {10.1134/S1063772915070021}, \href
  {http://adsabs.harvard.edu/abs/2015ARep...59..747A} {59, 747}

\bibitem[\protect\citeauthoryear{{Akimkin}, {Zhukovska}, {Wiebe}, {Semenov},
  {Pavlyuchenkov}, {Vasyunin}, {Birnstiel}  \& {Henning}}{{Akimkin}
  et~al.}{2013}]{Akimkin2013}
{Akimkin} V.,  {Zhukovska} S.,  {Wiebe} D.,  {Semenov} D.,  {Pavlyuchenkov} Y.,
   {Vasyunin} A.,  {Birnstiel} T.,   {Henning} T.,  2013, \mn@doi [\apj]
  {10.1088/0004-637X/766/1/8}, \href
  {http://adsabs.harvard.edu/abs/2013ApJ...766....8A} {766, 8}

\bibitem[\protect\citeauthoryear{{Andrews}, {Wilner}, {Espaillat}, {Hughes},
  {Dullemond}, {McClure}, {Qi}  \& {Brown}}{{Andrews}
  et~al.}{2011}]{Andrews2011}
{Andrews} S.~M.,  {Wilner} D.~J.,  {Espaillat} C.,  {Hughes} A.~M.,
  {Dullemond} C.~P.,  {McClure} M.~K.,  {Qi} C.,   {Brown} J.~M.,  2011,
  \mn@doi [\apj] {10.1088/0004-637X/732/1/42}, \href
  {http://adsabs.harvard.edu/abs/2011ApJ...732...42A} {732, 42}

\bibitem[\protect\citeauthoryear{{Bae} \& {Zhu}}{{Bae} \&
  {Zhu}}{2017}]{Bae2017}
{Bae} J.,  {Zhu} Z.,  2017, preprint, \href
  {http://adsabs.harvard.edu/abs/2017arXiv171108161B} {} (\mn@eprint {arXiv}
  {1711.08161})

\bibitem[\protect\citeauthoryear{{Balbus} \& {Hawley}}{{Balbus} \&
  {Hawley}}{1992}]{Balbus1992}
{Balbus} S.~A.,  {Hawley} J.~F.,  1992, \mn@doi [\apj] {10.1086/172022}, \href
  {http://adsabs.harvard.edu/abs/1992ApJ...400..610B} {400, 610}

\bibitem[\protect\citeauthoryear{{Benisty} et~al.,}{{Benisty}
  et~al.}{2015}]{Benisty2015}
{Benisty} M.,  et~al., 2015, \mn@doi [\aap] {10.1051/0004-6361/201526011},
  \href {http://adsabs.harvard.edu/abs/2015A%26A...578L...6B} {578, L6}

\bibitem[\protect\citeauthoryear{{Boley}}{{Boley}}{2009}]{Boley2009}
{Boley} A.~C.,  2009, \mn@doi [\apjl] {10.1088/0004-637X/695/1/L53}, \href
  {http://adsabs.harvard.edu/abs/2009ApJ...695L..53B} {695, L53}

\bibitem[\protect\citeauthoryear{{Boss}}{{Boss}}{1997}]{Boss97}
{Boss} A.~P.,  1997, \mn@doi [Science] {10.1126/science.276.5320.1836}, \href
  {http://adsabs.harvard.edu/abs/1997Sci...276.1836B} {276, 1836}

\bibitem[\protect\citeauthoryear{{Boss}}{{Boss}}{2017}]{Boss2017}
{Boss} A.~P.,  2017, \mn@doi [\apj] {10.3847/1538-4357/836/1/53}, \href
  {http://adsabs.harvard.edu/abs/2017ApJ...836...53B} {836, 53}

\bibitem[\protect\citeauthoryear{{Cameron}}{{Cameron}}{1973}]{Cameron1973}
{Cameron} A.~G.~W.,  1973, \mn@doi [\icarus] {10.1016/0019-1035(73)90153-X},
  \href {http://adsabs.harvard.edu/abs/1973Icar...18..407C} {18, 407}

\bibitem[\protect\citeauthoryear{{Chatterjee} \& {Tan}}{{Chatterjee} \&
  {Tan}}{2014}]{Chatt2014}
{Chatterjee} S.,  {Tan} J.~C.,  2014, \mn@doi [\apj]
  {10.1088/0004-637X/780/1/53}, \href
  {http://adsabs.harvard.edu/abs/2014ApJ...780...53C} {780, 53}

\bibitem[\protect\citeauthoryear{{Collin} \& {Zahn}}{{Collin} \&
  {Zahn}}{2008}]{Collin2008}
{Collin} S.,  {Zahn} J.-P.,  2008, \mn@doi [\aap] {10.1051/0004-6361:20078191},
  \href {http://adsabs.harvard.edu/abs/2008A%26A...477..419C} {477, 419}

\bibitem[\protect\citeauthoryear{{Coradini}, {Magni}  \& {Federico}}{{Coradini}
  et~al.}{1981}]{cora}
{Coradini} A.,  {Magni} G.,   {Federico} C.,  1981, A\& A, \href
  {http://adsabs.harvard.edu/abs/1981A%26A....98..173C} {98, 173}

\bibitem[\protect\citeauthoryear{{D'Angelo}, {Durisen}  \&
  {Lissauer}}{{D'Angelo} et~al.}{2010}]{D'Angelo2010}
{D'Angelo} G.,  {Durisen} R.~H.,   {Lissauer} J.~J.,  2010, {Giant Planet
  Formation}.
pp 319--346

\bibitem[\protect\citeauthoryear{{Dipierro} et~al.,}{{Dipierro}
  et~al.}{2018}]{Dipierro18}
{Dipierro} G.,  et~al., 2018, \mn@doi [\mnras] {10.1093/mnras/sty181}, \href
  {http://adsabs.harvard.edu/abs/2018MNRAS.475.5296D} {475, 5296}

\bibitem[\protect\citeauthoryear{{Dong}, {Hall}, {Rice}  \& {Chiang}}{{Dong}
  et~al.}{2015}]{Dong2015}
{Dong} R.,  {Hall} C.,  {Rice} K.,   {Chiang} E.,  2015, \mn@doi [\apjl]
  {10.1088/2041-8205/812/2/L32}, \href
  {http://adsabs.harvard.edu/abs/2015ApJ...812L..32D} {812, L32}

\bibitem[\protect\citeauthoryear{{Durisen}, {Boss}, {Mayer}, {Nelson}, {Quinn}
  \& {Rice}}{{Durisen} et~al.}{2007}]{Durisen2007}
{Durisen} R.~H.,  {Boss} A.~P.,  {Mayer} L.,  {Nelson} A.~F.,  {Quinn} T.,
  {Rice} W.~K.~M.,  2007, Protostars and Planets V, \href
  {http://adsabs.harvard.edu/abs/2007prpl.conf..607D} {pp 607--622}

\bibitem[\protect\citeauthoryear{{Fuchs}}{{Fuchs}}{2001}]{Fuchs2001}
{Fuchs} B.,  2001, \mn@doi [\aap] {10.1051/0004-6361:20000562}, \href
  {https://ui.adsabs.harvard.edu/abs/2001A&A...368..107F} {368, 107}

\bibitem[\protect\citeauthoryear{{Gammie}}{{Gammie}}{2001}]{Gammie2001}
{Gammie} C.~F.,  2001, \mn@doi [\apj] {10.1086/320631}, \href
  {http://adsabs.harvard.edu/abs/2001ApJ...553..174G} {553, 174}

\bibitem[\protect\citeauthoryear{{Garufi} et~al.,}{{Garufi}
  et~al.}{2013}]{Garufi2013}
{Garufi} A.,  et~al., 2013, \mn@doi [\aap] {10.1051/0004-6361/201322429}, \href
  {http://adsabs.harvard.edu/abs/2013A%26A...560A.105G} {560, A105}

\bibitem[\protect\citeauthoryear{{Ghosh} \& {Jog}}{{Ghosh} \&
  {Jog}}{2018}]{Ghosh18}
{Ghosh} S.,  {Jog} C.~J.,  2018, \mn@doi [\aap] {10.1051/0004-6361/201832988},
  \href {https://ui.adsabs.harvard.edu/abs/2018A&A...617A..47G} {617, A47}

\bibitem[\protect\citeauthoryear{{Goldbaum}, {Krumholz}  \&
  {Forbes}}{{Goldbaum} et~al.}{2016}]{Goldbaum2016}
{Goldbaum} N.~J.,  {Krumholz} M.~R.,   {Forbes} J.~C.,  2016, \mn@doi [\apj]
  {10.3847/0004-637X/827/1/28}, \href
  {http://adsabs.harvard.edu/abs/2016ApJ...827...28G} {827, 28}

\bibitem[\protect\citeauthoryear{{Goldreich} \& {Lynden-Bell}}{{Goldreich} \&
  {Lynden-Bell}}{1965}]{gold}
{Goldreich} P.,  {Lynden-Bell} D.,  1965, MNRAS, \href
  {http://adsabs.harvard.edu/abs/1965MNRAS.130..125G} {130, 125}

\bibitem[\protect\citeauthoryear{{Goldreich} \& {Ward}}{{Goldreich} \&
  {Ward}}{1973}]{Goldreich1973}
{Goldreich} P.,  {Ward} W.~R.,  1973, \mn@doi [\apj] {10.1086/152291}, \href
  {http://adsabs.harvard.edu/abs/1973ApJ...183.1051G} {183, 1051}

\bibitem[\protect\citeauthoryear{{Grady} et~al.,}{{Grady}
  et~al.}{2013}]{Grady2013}
{Grady} C.~A.,  et~al., 2013, \mn@doi [\apj] {10.1088/0004-637X/762/1/48},
  \href {http://adsabs.harvard.edu/abs/2013ApJ...762...48G} {762, 48}

\bibitem[\protect\citeauthoryear{{Hayashi}}{{Hayashi}}{1981}]{hayashi81}
{Hayashi} C.,  1981, \mn@doi [Progress of Theoretical Physics Supplement]
  {10.1143/PTPS.70.35}, \href
  {http://adsabs.harvard.edu/abs/1981PThPS..70...35H} {70, 35}

\bibitem[\protect\citeauthoryear{{Hayashi}, {Nakazawa}  \& {Adachi}}{{Hayashi}
  et~al.}{1977}]{Hayashi1977}
{Hayashi} C.,  {Nakazawa} K.,   {Adachi} I.,  1977, \pasj, \href
  {http://adsabs.harvard.edu/abs/1977PASJ...29..163H} {29, 163}

\bibitem[\protect\citeauthoryear{{Helled} et~al.,}{{Helled}
  et~al.}{2014}]{Helled2014}
{Helled} R.,  et~al., 2014, \mn@doi [Protostars and Planets VI]
  {10.2458/azu_uapress_9780816531240-ch028}, \href
  {http://adsabs.harvard.edu/abs/2014prpl.conf..643H} {pp 643--665}

\bibitem[\protect\citeauthoryear{{Hendler} et~al.,}{{Hendler}
  et~al.}{2017}]{Hendler2017}
{Hendler} N.~P.,  et~al., 2017, preprint, \href
  {http://adsabs.harvard.edu/abs/2017arXiv171109933H} {} (\mn@eprint {arXiv}
  {1711.09933})

\bibitem[\protect\citeauthoryear{{Ivlev}, {Akimkin}  \& {Caselli}}{{Ivlev}
  et~al.}{2016}]{Ivlev2016}
{Ivlev} A.~V.,  {Akimkin} V.~V.,   {Caselli} P.,  2016, \mn@doi [\apj]
  {10.3847/1538-4357/833/1/92}, \href
  {http://adsabs.harvard.edu/abs/2016ApJ...833...92I} {833, 92}

\bibitem[\protect\citeauthoryear{{Jacquet}, {Balbus}  \& {Latter}}{{Jacquet}
  et~al.}{2011}]{Jac}
{Jacquet} E.,  {Balbus} S.,   {Latter} H.,  2011, \mn@doi [MNRAS]
  {10.1111/j.1365-2966.2011.18971.x}, \href
  {http://adsabs.harvard.edu/abs/2011MNRAS.415.3591J} {415, 3591}

\bibitem[\protect\citeauthoryear{{Jog}}{{Jog}}{1992}]{Jog1992}
{Jog} C.~J.,  1992, \mn@doi [\apj] {10.1086/171289}, \href
  {http://adsabs.harvard.edu/abs/1992ApJ...390..378J} {390, 378}

\bibitem[\protect\citeauthoryear{{Juh{\'a}sz} \& {Rosotti}}{{Juh{\'a}sz} \&
  {Rosotti}}{2018}]{Juhasz2018}
{Juh{\'a}sz} A.,  {Rosotti} G.~P.,  2018, \mn@doi [\mnras]
  {10.1093/mnrasl/slx182}, \href
  {http://adsabs.harvard.edu/abs/2018MNRAS.474L..32J} {474, L32}

\bibitem[\protect\citeauthoryear{{Julian} \& {Toomre}}{{Julian} \&
  {Toomre}}{1966}]{Julian1966}
{Julian} W.~H.,  {Toomre} A.,  1966, \mn@doi [\apj] {10.1086/148957}, \href
  {http://adsabs.harvard.edu/abs/1966ApJ...146..810J} {146, 810}

\bibitem[\protect\citeauthoryear{{Kratter}, {Matzner}  \& {Krumholz}}{{Kratter}
  et~al.}{2008}]{Kratter2008}
{Kratter} K.~M.,  {Matzner} C.~D.,   {Krumholz} M.~R.,  2008, \mn@doi [\apj]
  {10.1086/587543}, \href {http://adsabs.harvard.edu/abs/2008ApJ...681..375K}
  {681, 375}

\bibitem[\protect\citeauthoryear{{Krumholz} \& {Burkert}}{{Krumholz} \&
  {Burkert}}{2010}]{Krumholz2010}
{Krumholz} M.,  {Burkert} A.,  2010, \mn@doi [\apj]
  {10.1088/0004-637X/724/2/895}, \href
  {http://adsabs.harvard.edu/abs/2010ApJ...724..895K} {724, 895}

\bibitem[\protect\citeauthoryear{{Latter} \& {Rosca}}{{Latter} \&
  {Rosca}}{2017}]{Latter17}
{Latter} H.~N.,  {Rosca} R.,  2017, \mn@doi [\mnras] {10.1093/mnras/stw2455},
  \href {http://adsabs.harvard.edu/abs/2017MNRAS.464.1923L} {464, 1923}

\bibitem[\protect\citeauthoryear{{Lee}}{{Lee}}{2016}]{Lee2016}
{Lee} W.-K.,  2016, \mn@doi [\apj] {10.3847/0004-637X/832/2/166}, \href
  {http://adsabs.harvard.edu/abs/2016ApJ...832..166L} {832, 166}

\bibitem[\protect\citeauthoryear{{Lin} \& {Shu}}{{Lin} \&
  {Shu}}{1964}]{Lin1964}
{Lin} C.~C.,  {Shu} F.~H.,  1964, \mn@doi [\apj] {10.1086/147955}, \href
  {http://adsabs.harvard.edu/abs/1964ApJ...140..646L} {140, 646}

\bibitem[\protect\citeauthoryear{{Lissauer}}{{Lissauer}}{1993}]{Lissauer1993}
{Lissauer} J.~J.,  1993, \mn@doi [\araa] {10.1146/annurev.aa.31.090193.001021},
  \href {http://adsabs.harvard.edu/abs/1993ARA%26A..31..129L} {31, 129}

\bibitem[\protect\citeauthoryear{{Lissauer} \& {Stevenson}}{{Lissauer} \&
  {Stevenson}}{2007}]{Lissauer2007}
{Lissauer} J.~J.,  {Stevenson} D.~J.,  2007, Protostars and Planets V, \href
  {http://adsabs.harvard.edu/abs/2007prpl.conf..591L} {pp 591--606}

\bibitem[\protect\citeauthoryear{{Loomis}, {{\"O}berg}, {Andrews}  \&
  {MacGregor}}{{Loomis} et~al.}{2017}]{Loomis2017}
{Loomis} R.~A.,  {{\"O}berg} K.~I.,  {Andrews} S.~M.,   {MacGregor} M.~A.,
  2017, \mn@doi [\apj] {10.3847/1538-4357/aa6c63}, \href
  {http://adsabs.harvard.edu/abs/2017ApJ...840...23L} {840, 23}

\bibitem[\protect\citeauthoryear{{Mamatsashvili}, {Chagelishvili}, {Bodo}  \&
  {Rossi}}{{Mamatsashvili} et~al.}{2013}]{Mamats2013}
{Mamatsashvili} G.~R.,  {Chagelishvili} G.~D.,  {Bodo} G.,   {Rossi} P.,  2013,
  \mn@doi [\mnras] {10.1093/mnras/stt1470}, \href
  {http://adsabs.harvard.edu/abs/2013MNRAS.435.2552M} {435, 2552}

\bibitem[\protect\citeauthoryear{{Matzner} \& {Levin}}{{Matzner} \&
  {Levin}}{2005}]{Matzner2005}
{Matzner} C.~D.,  {Levin} Y.,  2005, \mn@doi [\apj] {10.1086/430813}, \href
  {http://adsabs.harvard.edu/abs/2005ApJ...628..817M} {628, 817}

\bibitem[\protect\citeauthoryear{{Mayama} et~al.,}{{Mayama}
  et~al.}{2012}]{Mayama2012}
{Mayama} S.,  et~al., 2012, \mn@doi [\apjl] {10.1088/2041-8205/760/2/L26},
  \href {http://adsabs.harvard.edu/abs/2012ApJ...760L..26M} {760, L26}

\bibitem[\protect\citeauthoryear{{Mej{\'{\i}}a}, {Durisen}, {Pickett}  \&
  {Cai}}{{Mej{\'{\i}}a} et~al.}{2005}]{Mejia2005}
{Mej{\'{\i}}a} A.~C.,  {Durisen} R.~H.,  {Pickett} M.~K.,   {Cai} K.,  2005,
  \mn@doi [\apj] {10.1086/426707}, \href
  {http://adsabs.harvard.edu/abs/2005ApJ...619.1098M} {619, 1098}

\bibitem[\protect\citeauthoryear{{Michikoshi} \& {Kokubo}}{{Michikoshi} \&
  {Kokubo}}{2016}]{Mich2016}
{Michikoshi} S.,  {Kokubo} E.,  2016, \mn@doi [\apj]
  {10.3847/0004-637X/823/2/121}, \href
  {https://ui.adsabs.harvard.edu/abs/2016ApJ...823..121M} {823, 121}

\bibitem[\protect\citeauthoryear{{Michikoshi}, {Kokubo}  \&
  {Inutsuka}}{{Michikoshi} et~al.}{2012}]{Mich}
{Michikoshi} S.,  {Kokubo} E.,   {Inutsuka} S.-i.,  2012, \mn@doi [ApJ]
  {10.1088/0004-637X/746/1/35}, \href
  {http://adsabs.harvard.edu/abs/2012ApJ...746...35M} {746, 35}

\bibitem[\protect\citeauthoryear{{Miyake}, {Suzuki}  \& {Inutsuka}}{{Miyake}
  et~al.}{2016}]{Miyake2016}
{Miyake} T.,  {Suzuki} T.~K.,   {Inutsuka} S.-i.,  2016, \mn@doi [\apj]
  {10.3847/0004-637X/821/1/3}, \href
  {http://adsabs.harvard.edu/abs/2016ApJ...821....3M} {821, 3}

\bibitem[\protect\citeauthoryear{{Mizuno}}{{Mizuno}}{1980}]{Mizuno1980}
{Mizuno} H.,  1980, \mn@doi [Progress of Theoretical Physics]
  {10.1143/PTP.64.544}, \href
  {http://adsabs.harvard.edu/abs/1980PThPh..64..544M} {64, 544}

\bibitem[\protect\citeauthoryear{{Muto} et~al.,}{{Muto}
  et~al.}{2012}]{Muto2012}
{Muto} T.,  et~al., 2012, \mn@doi [\apjl] {10.1088/2041-8205/748/2/L22}, \href
  {http://adsabs.harvard.edu/abs/2012ApJ...748L..22M} {748, L22}

\bibitem[\protect\citeauthoryear{{Natta}, {Testi}, {Calvet}, {Henning},
  {Waters}  \& {Wilner}}{{Natta} et~al.}{2007}]{Natta2007}
{Natta} A.,  {Testi} L.,  {Calvet} N.,  {Henning} T.,  {Waters} R.,   {Wilner}
  D.,  2007, Protostars and Planets V, \href
  {http://adsabs.harvard.edu/abs/2007prpl.conf..767N} {pp 767--781}

\bibitem[\protect\citeauthoryear{{Nayakshin}}{{Nayakshin}}{2010}]{Nayakshin2010}
{Nayakshin} S.,  2010, \mn@doi [\mnras] {10.1111/j.1745-3933.2010.00923.x},
  \href {http://adsabs.harvard.edu/abs/2010MNRAS.408L..36N} {408, L36}

\bibitem[\protect\citeauthoryear{{Nixon}, {King}  \& {Pringle}}{{Nixon}
  et~al.}{2018}]{Nixon2018}
{Nixon} C.~J.,  {King} A.~R.,   {Pringle} J.~E.,  2018, \mn@doi [\mnras]
  {10.1093/mnras/sty593}, \href
  {http://adsabs.harvard.edu/abs/2018MNRAS.477.3273N} {477, 3273}

\bibitem[\protect\citeauthoryear{{Noh}, {Vishniac}  \& {Cochran}}{{Noh}
  et~al.}{1991}]{noh91}
{Noh} H.,  {Vishniac} E.~T.,   {Cochran} W.~D.,  1991, \mn@doi [ApJ]
  {10.1086/170795}, \href {http://adsabs.harvard.edu/abs/1991ApJ...383..372N}
  {383, 372}

\bibitem[\protect\citeauthoryear{{Ogilvie} \& {Lubow}}{{Ogilvie} \&
  {Lubow}}{2002}]{Ogilvie2002}
{Ogilvie} G.~I.,  {Lubow} S.~H.,  2002, \mn@doi [\mnras]
  {10.1046/j.1365-8711.2002.05148.x}, \href
  {http://adsabs.harvard.edu/abs/2002MNRAS.330..950O} {330, 950}

\bibitem[\protect\citeauthoryear{{Okuzumi}, {Tanaka}, {Takeuchi}  \&
  {Sakagami}}{{Okuzumi} et~al.}{2011}]{Okuzumi2011}
{Okuzumi} S.,  {Tanaka} H.,  {Takeuchi} T.,   {Sakagami} M.-a.,  2011, \mn@doi
  [\apj] {10.1088/0004-637X/731/2/96}, \href
  {http://adsabs.harvard.edu/abs/2011ApJ...731...96O} {731, 96}

\bibitem[\protect\citeauthoryear{{Okuzumi}, {Momose}, {Sirono}, {Kobayashi}  \&
  {Tanaka}}{{Okuzumi} et~al.}{2016}]{Oku16}
{Okuzumi} S.,  {Momose} M.,  {Sirono} S.-i.,  {Kobayashi} H.,   {Tanaka} H.,
  2016, \mn@doi [\apj] {10.3847/0004-637X/821/2/82}, \href
  {http://adsabs.harvard.edu/abs/2016ApJ...821...82O} {821, 82}

\bibitem[\protect\citeauthoryear{{Pollack}, {Hubickyj}, {Bodenheimer},
  {Lissauer}, {Podolak}  \& {Greenzweig}}{{Pollack} et~al.}{1996}]{Pollack1996}
{Pollack} J.~B.,  {Hubickyj} O.,  {Bodenheimer} P.,  {Lissauer} J.~J.,
  {Podolak} M.,   {Greenzweig} Y.,  1996, \mn@doi [\icarus]
  {10.1006/icar.1996.0190}, \href
  {http://adsabs.harvard.edu/abs/1996Icar..124...62P} {124, 62}

\bibitem[\protect\citeauthoryear{{Rab}, {G{\"u}del}, {Woitke}, {Kamp}, {Thi},
  {Min}, {Aresu}  \& {Meijerink}}{{Rab} et~al.}{2017}]{Rab2017}
{Rab} C.,  {G{\"u}del} M.,  {Woitke} P.,  {Kamp} I.,  {Thi} W.-F.,  {Min} M.,
  {Aresu} G.,   {Meijerink} R.,  2017, preprint, \href
  {http://adsabs.harvard.edu/abs/2017arXiv171107249R} {} (\mn@eprint {arXiv}
  {1711.07249})

\bibitem[\protect\citeauthoryear{{Rafikov}}{{Rafikov}}{2005}]{Rafikov2005}
{Rafikov} R.~R.,  2005, \mn@doi [\apjl] {10.1086/428899}, \href
  {http://adsabs.harvard.edu/abs/2005ApJ...621L..69R} {621, L69}

\bibitem[\protect\citeauthoryear{{Rafikov}}{{Rafikov}}{2009}]{Rafikov2009}
{Rafikov} R.~R.,  2009, \mn@doi [\apj] {10.1088/0004-637X/704/1/281}, \href
  {http://adsabs.harvard.edu/abs/2009ApJ...704..281R} {704, 281}

\bibitem[\protect\citeauthoryear{{Rice}, {Armitage}, {Bate}  \&
  {Bonnell}}{{Rice} et~al.}{2003}]{Rice2003}
{Rice} W.~K.~M.,  {Armitage} P.~J.,  {Bate} M.~R.,   {Bonnell} I.~A.,  2003,
  \mn@doi [\mnras] {10.1046/j.1365-8711.2003.06253.x}, \href
  {http://adsabs.harvard.edu/abs/2003MNRAS.339.1025R} {339, 1025}

\bibitem[\protect\citeauthoryear{{Romeo} \& {Agertz}}{{Romeo} \&
  {Agertz}}{2014}]{Romeo2014}
{Romeo} A.~B.,  {Agertz} O.,  2014, \mn@doi [\mnras] {10.1093/mnras/stu954},
  \href {http://adsabs.harvard.edu/abs/2014MNRAS.442.1230R} {442, 1230}

\bibitem[\protect\citeauthoryear{{Safronov}}{{Safronov}}{1972}]{Safronov1972}
{Safronov} V.~S.,  1972, {Evolution of the protoplanetary cloud and formation
  of the earth and planets.}

\bibitem[\protect\citeauthoryear{{Sekiya}}{{Sekiya}}{1983}]{sekiya83}
{Sekiya} M.,  1983, \mn@doi [Progress of Theoretical Physics]
  {10.1143/PTP.69.1116}, \href
  {http://adsabs.harvard.edu/abs/1983PThPh..69.1116S} {69, 1116}

\bibitem[\protect\citeauthoryear{{Shariff} \& {Cuzzi}}{{Shariff} \&
  {Cuzzi}}{2011}]{Cuzzi}
{Shariff} K.,  {Cuzzi} J.~N.,  2011, \mn@doi [ApJ]
  {10.1088/0004-637X/738/1/73}, \href
  {http://adsabs.harvard.edu/abs/2011ApJ...738...73S} {738, 73}

\bibitem[\protect\citeauthoryear{{Shu}}{{Shu}}{1984}]{Shu1984}
{Shu} F.~H.,  1984, in {Greenberg} R.,  {Brahic} A.,  eds, IAU Colloq. 75:
  Planetary Rings. pp 513--561

\bibitem[\protect\citeauthoryear{{Stevenson}}{{Stevenson}}{1982}]{Stevenson1982}
{Stevenson} D.~J.,  1982, \mn@doi [\planss] {10.1016/0032-0633(82)90108-8},
  \href {http://adsabs.harvard.edu/abs/1982P%26SS...30..755S} {30, 755}

\bibitem[\protect\citeauthoryear{{Suriano}, {Li}, {Krasnopolsky}  \&
  {Shang}}{{Suriano} et~al.}{2018}]{Suriano18}
{Suriano} S.~S.,  {Li} Z.-Y.,  {Krasnopolsky} R.,   {Shang} H.,  2018, \mn@doi
  [\mnras] {10.1093/mnras/sty717}, \href
  {http://adsabs.harvard.edu/abs/2018MNRAS.477.1239S} {477, 1239}

\bibitem[\protect\citeauthoryear{{Takahashi} \& {Inutsuka}}{{Takahashi} \&
  {Inutsuka}}{2014}]{Taka}
{Takahashi} S.~Z.,  {Inutsuka} S.-i.,  2014, \mn@doi [ApJ]
  {10.1088/0004-637X/794/1/55}, \href
  {http://adsabs.harvard.edu/abs/2014ApJ...794...55T} {794, 55}

\bibitem[\protect\citeauthoryear{{Takahashi} \& {Inutsuka}}{{Takahashi} \&
  {Inutsuka}}{2016}]{TaInu2016}
{Takahashi} S.~Z.,  {Inutsuka} S.-i.,  2016, \mn@doi [\aj]
  {10.3847/0004-6256/152/6/184}, \href
  {http://adsabs.harvard.edu/abs/2016AJ....152..184T} {152, 184}

\bibitem[\protect\citeauthoryear{{Takahashi}, {Tsukamoto}  \&
  {Inutsuka}}{{Takahashi} et~al.}{2016}]{Takahashi2016}
{Takahashi} S.~Z.,  {Tsukamoto} Y.,   {Inutsuka} S.,  2016, \mn@doi [\mnras]
  {10.1093/mnras/stw557}, \href
  {http://adsabs.harvard.edu/abs/2016MNRAS.458.3597T} {458, 3597}

\bibitem[\protect\citeauthoryear{{Tobin} et~al.,}{{Tobin}
  et~al.}{2016}]{Tobin2016}
{Tobin} J.~J.,  et~al., 2016, \mn@doi [\nat] {10.1038/nature20094}, \href
  {http://adsabs.harvard.edu/abs/2016Natur.538..483T} {538, 483}

\bibitem[\protect\citeauthoryear{{Tomida}, {Machida}, {Hosokawa}, {Sakurai}  \&
  {Lin}}{{Tomida} et~al.}{2017}]{Tomida2017}
{Tomida} K.,  {Machida} M.~N.,  {Hosokawa} T.,  {Sakurai} Y.,   {Lin} C.~H.,
  2017, \mn@doi [\apjl] {10.3847/2041-8213/835/1/L11}, \href
  {http://adsabs.harvard.edu/abs/2017ApJ...835L..11T} {835, L11}

\bibitem[\protect\citeauthoryear{{Tominaga}, {Inutsuka}  \&
  {Takahashi}}{{Tominaga} et~al.}{2018}]{Tominaga2018}
{Tominaga} R.~T.,  {Inutsuka} S.-i.,   {Takahashi} S.~Z.,  2018, \mn@doi
  [\pasj] {10.1093/pasj/psx143}, \href
  {http://adsabs.harvard.edu/abs/2018PASJ...70....3T} {70, 3}

\bibitem[\protect\citeauthoryear{{Toomre}}{{Toomre}}{1964}]{Toomre1964}
{Toomre} A.,  1964, \mn@doi [\apj] {10.1086/147861}, \href
  {http://adsabs.harvard.edu/abs/1964ApJ...139.1217T} {139, 1217}

\bibitem[\protect\citeauthoryear{{Toomre}}{{Toomre}}{1981}]{Toomre1981}
{Toomre} A.,  1981, in {Fall} S.~M.,  {Lynden-Bell} D.,  eds, Structure and
  Evolution of Normal Galaxies. pp 111--136

\bibitem[\protect\citeauthoryear{{Tsukamoto}, {Takahashi}, {Machida}  \&
  {Inutsuka}}{{Tsukamoto} et~al.}{2015}]{Tsukamoto2015}
{Tsukamoto} Y.,  {Takahashi} S.~Z.,  {Machida} M.~N.,   {Inutsuka} S.,  2015,
  \mn@doi [\mnras] {10.1093/mnras/stu2160}, \href
  {http://adsabs.harvard.edu/abs/2015MNRAS.446.1175T} {446, 1175}

\bibitem[\protect\citeauthoryear{{Vandervoort}}{{Vandervoort}}{1970}]{Vandervoort1970}
{Vandervoort} P.~O.,  1970, \mn@doi [\apj] {10.1086/150514}, \href
  {http://adsabs.harvard.edu/abs/1970ApJ...161...87V} {161, 87}

\bibitem[\protect\citeauthoryear{{Vasyunin}, {Wiebe}, {Birnstiel}, {Zhukovska},
  {Henning}  \& {Dullemond}}{{Vasyunin} et~al.}{2011}]{Vasyunin2011}
{Vasyunin} A.~I.,  {Wiebe} D.~S.,  {Birnstiel} T.,  {Zhukovska} S.,  {Henning}
  T.,   {Dullemond} C.~P.,  2011, \mn@doi [\apj] {10.1088/0004-637X/727/2/76},
  \href {http://adsabs.harvard.edu/abs/2011ApJ...727...76V} {727, 76}

\bibitem[\protect\citeauthoryear{{Wagner}, {Apai}, {Kasper}  \&
  {Robberto}}{{Wagner} et~al.}{2015}]{Wagner2015}
{Wagner} K.,  {Apai} D.,  {Kasper} M.,   {Robberto} M.,  2015, \mn@doi [\apjl]
  {10.1088/2041-8205/813/1/L2}, \href
  {http://adsabs.harvard.edu/abs/2015ApJ...813L...2W} {813, L2}

\bibitem[\protect\citeauthoryear{{Woitke} et~al.,}{{Woitke}
  et~al.}{2016}]{Woitke2016}
{Woitke} P.,  et~al., 2016, \mn@doi [\aap] {10.1051/0004-6361/201526538}, \href
  {http://adsabs.harvard.edu/abs/2016A%26A...586A.103W} {586, A103}

\bibitem[\protect\citeauthoryear{{Yen}, {Liu}, {Gu}, {Hirano}, {Lee},
  {Puspitaningrum}  \& {Takakuwa}}{{Yen} et~al.}{2016}]{Yen16}
{Yen} H.-W.,  {Liu} H.~B.,  {Gu} P.-G.,  {Hirano} N.,  {Lee} C.-F.,
  {Puspitaningrum} E.,   {Takakuwa} S.,  2016, \mn@doi [\apjl]
  {10.3847/2041-8205/820/2/L25}, \href
  {http://adsabs.harvard.edu/abs/2016ApJ...820L..25Y} {820, L25}

\bibitem[\protect\citeauthoryear{{Youdin}}{{Youdin}}{2011}]{youdin2011}
{Youdin} A.~N.,  2011, \mn@doi [ApJ] {10.1088/0004-637X/731/2/99}, \href
  {http://adsabs.harvard.edu/abs/2011ApJ...731...99Y} {731, 99}

\bibitem[\protect\citeauthoryear{{Youdin} \& {Goodman}}{{Youdin} \&
  {Goodman}}{2005}]{youdin2005}
{Youdin} A.~N.,  {Goodman} J.,  2005, \mn@doi [ApJ] {10.1086/426895}, \href
  {http://adsabs.harvard.edu/abs/2005ApJ...620..459Y} {620, 459}

\bibitem[\protect\citeauthoryear{{Youdin} \& {Lithwick}}{{Youdin} \&
  {Lithwick}}{2007}]{youdin2007}
{Youdin} A.~N.,  {Lithwick} Y.,  2007, \mn@doi [Icarus]
  {10.1016/j.icarus.2007.07.012}, \href
  {http://adsabs.harvard.edu/abs/2007Icar..192..588Y} {192, 588}

\bibitem[\protect\citeauthoryear{{Zhu}, {Dong}, {Stone}  \& {Rafikov}}{{Zhu}
  et~al.}{2015}]{Zhu2015}
{Zhu} Z.,  {Dong} R.,  {Stone} J.~M.,   {Rafikov} R.~R.,  2015, \mn@doi [\apj]
  {10.1088/0004-637X/813/2/88}, \href
  {http://adsabs.harvard.edu/abs/2015ApJ...813...88Z} {813, 88}

\bibitem[\protect\citeauthoryear{{van Terwisga} et~al.,}{{van Terwisga}
  et~al.}{2018}]{van2018}
{van Terwisga} S.~E.,  et~al., 2018, preprint, \href
  {http://adsabs.harvard.edu/abs/2018arXiv180503221V} {} (\mn@eprint {arXiv}
  {1805.03221})

\makeatother
\end{thebibliography}




\bsp	
\label{lastpage}
\end{document}